\documentclass[a4paper,longauth]{emulateapj}
\usepackage{apjfonts}
\usepackage{graphicx,subfigure}
\usepackage{epsfig}
\usepackage{epsf}
\usepackage{natbib}
\usepackage{multirow}
\usepackage{color}
\usepackage{hyperref}
\usepackage{hhline}
\usepackage{amsmath}
\newcommand{\be}{\begin{equation}}
\newcommand{\ee}{\end{equation}}
\newcommand{\ba}{\begin{array}}
\newcommand{\ea}{\end{array}}
\newcommand{\bea}{\begin{eqnarray}}
\newcommand{\eea}{\end{eqnarray}}
\newcommand{\mic}{\,{\rm \mu m} }

\shorttitle{Bayesian analysis of dust properties}
\shortauthors{M. Veneziani et al.}

\begin{document}   
\title{Bayesian method for the analysis of the dust emission in the Far-Infrared and Submillimeter }
%\subtitle{}

\author{M. Veneziani\altaffilmark{1,2}, 
F. Piacentini\altaffilmark{2}, 
A. Noriega-Crespo\altaffilmark{1}, 
S. Carey\altaffilmark{1}, 
R. Paladini\altaffilmark{1}, 
D. Paradis\altaffilmark{3,4}}

\email{marcella.veneziani@ipac.caltech.edu}

\altaffiltext{1}{Infrared Processing and Analysis Center, California Institute of Technology, Pasadena, CA 91125, USA}
\altaffiltext{2}{Dipartimento di Fisica, Universit\`a di Roma ``La Sapienza'', Rome, Italy}
\altaffiltext{3}{Universit\'e de Toulouse, UPS-OMP, IRAP, Toulouse, France }
\altaffiltext{4}{CNRS, IRAP, Toulouse, France}

\begin{abstract}
We present a method, based on Bayesian statistics, to fit the dust emission parameters in the far-infrared and
submillimeter wavelengths. The method estimates the dust temperature
and spectral emissivity index, plus their relationship, taking into account 
properly the statistical and systematic uncertainties.
We test it on three sets of simulated sources detectable by the Herschel Space Observatory in the PACS and SPIRE spectral bands (70-500 $\mic$), spanning over a wide range of dust temperatures. The simulated observations are a one-component Interstellar Medium, and two two-component sources, both warm (HII regions) and cold (cold clumps). 
We first define a procedure to identify the better model, then we recover the parameters of the model and measure their physical correlations by means of a Monte Carlo Markov Chain algorithm adopting multi-variate Gaussian priors. 
In this process we assess the reliability of the model recovery, and of parameters estimation. We conclude that the model and parameters are properly recovered only under certain circumstances, and that false models may be derived in some case.
We applied the method to a set of 91 starless cold clumps in an inter-arm region of the Galactic Plane with low star formation activity, observed by Herschel in the Hi-GAL survey. Our results are consistent with a temperature independent spectral index.

\end{abstract}

\section{Introduction}

Previous observations in the submillimeter (submm)
and far-infrared domains have enabled the investigation of 
the properties of dust in a variety of environments 
and at different angular resolutions. 
The brightness of radiation emitted from a  source in local 
thermal equilibrium has a continuum spectrum which can be expressed as 
$I_{\lambda} = \epsilon_{\lambda} B_{\lambda} (T)$,
where $\epsilon_{\lambda}$ is the wavelength dependent emissivity, and $B_{\lambda} (T)$ is the Planck 
function corresponding to a temperature $T$. In optically thin regions, the emissivity equals the optical depth $\tau_{\lambda}$. 
In optically thick regions the emissivity tends to 1. 
In the case of optically thin dust clouds~\citep{Desert90, Draine07}, 
the simplest models assume that the emissivity 
depends on the wavelength as
a power law, and the brightness spectrum is:
\begin{equation}
I_{\lambda}(\epsilon_0, \beta, T_d)  =  \epsilon_0 \left ( \frac{\lambda}{\lambda_0}
\right ) ^{-\beta} B_{\lambda}(T_d) 
\label{eq:Inu}
\end{equation}
where  T$\rm _d$ is the dust temperature, 
$\rm \epsilon_0$
is the emissivity at wavelength $\rm \lambda_0$, and $\rm \beta$ is
the emissivity spectral index, with a fiducial value $\beta = 2$.

More refined models describe the spectrum as due to a combination
of components different in temperature and nature. In particular, observational data 
evidence a flattening of the spectral index towards long wavelengths 
($\lambda > 500 \mu m$) which is well represented by the Multicomponent Dust Model 
described in~\citet{fds}.
The spectrum is then given by:
\begin{equation}
I_\lambda(\epsilon_1, \epsilon_2, T_1, T_2) = \epsilon_1 \left ( \frac{\lambda}{\lambda_0}
\right ) ^{-\beta_1} B_{\lambda}(T_1) + \epsilon_2 \left ( \frac{\lambda}{\lambda_0}
\right ) ^{-\beta_2} B_{\lambda}(T_2) 
\label{eq:Inu2}
\end{equation}
where $\epsilon_1$ and $\epsilon_2$ are the emissivities at $\lambda_0$,
$T_1$ and $T_2$ are the temperatures of the two dust components, and 
$\beta_1$ and $\beta_2$ are the spectral indices. 
Standard value for spectral indices can
be set to:
$\beta_1 = 1.67$, $\beta_2 = 2.70$, i.e. the best fit values in model 8 of~\citet{fds}.

More sophisticated models incorporate the effect of the
disordered internal structure of amorphous dust grains~\citep{Meny07}.
In this case, the emission is characterized by a single temperature 
and a spectral index which changes with temperature. Moreover, some level of anticorrelation 
between spectral index and temperature is
expected by laboratory experiments (see for example~\cite{coupeaud2011}). 
The emission can be then characterized by Eq.~(\ref{eq:Inu}) with the extra relation
\begin{equation}
\beta = A \left( \frac{T_d}{T_0}  \right)^\alpha
\label{eq:betat}
\end{equation}
where $T_0$ is a pivot temperature (we use $T_0 = 20$~K), and $A$ and $\alpha$ are parameters that 
characterize the inverse relation. 

Measurements from balloon based observatories and satellites indicates the existence of an anticorrelation
between $\beta$ and the T$_d$. 
The balloon-borne experiments PRONAOS
\citep{Dupac03} and ARCHEOPS \citep{Desert08} data sets evidenced an inverse
relationship between T$ _d$ and $\beta$ in a wide range of environments of the
interstellar medium (ISM) and in cold sources. In PRONAOS data, variations
in the spectral index were observed in the range 0.8-2.4 for dust temperatures
between 11 and 80 K, whereas ARCHEOPS data showed that the inverse
relationship is more pronounced with values of $\beta$ up to 4 for temperatures down to 7 K. 
Moreover, recently \cite{Veneziani10} highlighted a similar trend analyzing T$ _d$ and $\beta$ for eight high Galactic
latitudes clouds, by combining IRAS, DIRBE and WMAP data to BOOMERanG
observations. The $\beta$ values vary from 1 to 5 
in the temperature range 7-20 K, with a behavior similar to the
ARCHEOPS results. 
The analysis of the Herschel Hi-Gal key program \citep{Molinari10a,Molinari10b} has demonstrated this anticorrelation 
in the Interstellar Medium~\cite[ISM,][]{Paradis10} while the ATLAS found similar results
at high latitudes~\citep{Bracco11}. On the contrary,~\cite{Paladini12} finds the same inverse trend on HII regions but it 
might be generated from spurious effects. 

In order to test the hypothesis of the existence of a functional
dependency between the temperature and the spectral index, one has to
properly account for spurious correlations in their estimated
values. Due to the spectral shape in Eq.~(\ref{eq:Inu}), the effect of a
high value of T$_d$ can be mimicked, when holding the intensity fixed, by a
low value of $\beta$, and viceversa. There exists a statistical
degeneracy between these physical quantities that must be considered while investigating if an intrinsic physical correlation indeed exists.
\cite{Shetty09a,Shetty09b} report spurious inverse correlations due to a number of factors including noise in flux measurements coupling with the T$_d-\beta$ degeneracy and variations in temperature of overlapping sources along the sightline.

Because of the last Herschel and Planck submillimeter and millimeter data on dust, the topic of the $T_d-\beta$ spurious anticorrelation has been recently addressed~\cite[see for example][]{Juvela12a,Juvela12b,Kelly12} both on simulations and on real data. The originality of our paper is to treat statistical and systematic uncertainties in two different ways, both during the SED fitting and in the $T_d-\beta$ relationship estimate, in order to take into account their different statistical properties. 
In this paper we present a method based on Bayesian statistics, to discriminate between models in Eq.~(\ref{eq:Inu})-(\ref{eq:Inu2}) 
taking into account both the statistical and systematic errors of the input fluxes.
We also describe how to distinguish between a spurious and a physical 
anticorrelation between the spectral index and the temperature, 
and to estimate the $T_d-\beta$ inverse relationship by properly evaluating the degeneracy caused by the spectral shape and the 
flux uncertainties, both statistical and systematic, i.e. calibration uncertainty. 

We demonstrate the robustness of this method in the Herschel PACS and SPIRE spectral coverage, { $70\mu m < \lambda< 500 \mu m$}, and for a temperature range $10<T<50$K. 
We consider 3 cases: the diffuse ISM, warm sources (HII regions/YSOs), and cold sources (e.g. pre-stellar clumps).

This paper is organized as follows: Sec.~(\ref{sec:method}) presents the algorithm, after a brief introduction on the temperature-spectral index degeneracy and Bayesian statistics; Sec.~(\ref{sec:sources}) describes the simulated data set used in the analysis; Sec.~(\ref{sec:results}) and~\ref{sec:higal} report the results obtained by applying the method to the simulated observations and to cold clumps on the Galactic plane detected by the Herschel/Hi-GAL survey. Conclusions are drawn in Sec.~(\ref{sec:conclusions}). 

\section{Bayesian method for IR SEDs fitting}\label{sec:method}

{In this section, after a brief discussion of the temperature-spectral index degeneracy and of Bayesian statistics, we present our method to fit the SEDs taking into account both the statistical noise and the systematic uncertainties present in every dataset. The combination of Bayesian techniques and a good knowledge of the parameter distribution in the parameter space, allows to remove the spurious temperature spectral index anticorrelation, as well as the biases on the physical parameter estimates, generated both from random noise and systematic effects. }

\subsection{Bayesian treatment of statistical noise}\label{sec:noise}

{In order to study the effect of noise (i.e. statistical uncertainties) on one-component SED fitting, and in particular the $T_d-\beta$ degeneracy, we first simulate one ISM like and one HII region like source with a 20$\%$ gaussian random noise, and then fit for the physical parameters using in Eq.~(\ref{eq:Inu}) . 
Since we are interested in Herschel-like observations, we sample the SEDs in the six bands $\lambda=\left[70, 100, 160, 250, 350, 500\right]\mic$.
The simulated source temperatures are $\mathrm{T_{ISM}=15}$ K and ${T_{HII}=80}$K. 
The reason for chosing so different temperatures is shown in Fig. ~(\ref{fig:sed_nocorr}) where
the normalized SED of three sources at 15K, 40K and 80K and their sampling with the Herschel bands are reported. 
\begin{figure}[!t]
\begin{center}
\includegraphics[width=0.5\textwidth]{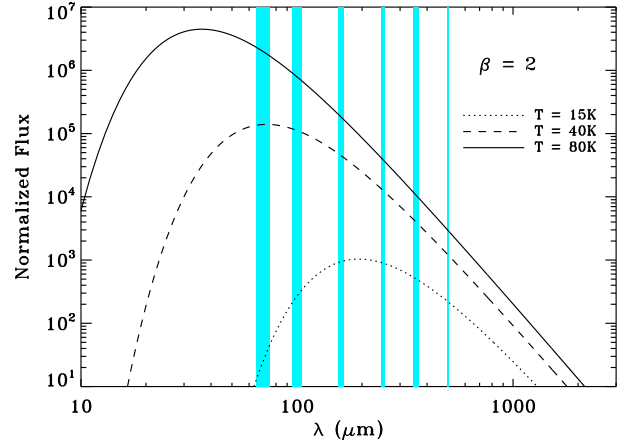}
\caption{Normalized SEDs for a 15K, 40K and 80K source and spectral index $\beta = 2$. The blue vertical lines identify the Herschel bands chosen to perform our analysis. The SED are then  well constrained at low temperatures (dotted line) and poorly constrained at high temperatures (solid line)
where Herschel sampling covers only the RJ part of the spectrum. }
\vspace{0.5cm}
\label{fig:sed_nocorr}
\end{center}
\end{figure}
In the Herschel bands the SED is well measured at low temperature (dotted line) while for warmer sources (dashed and solid lines) we can only sample the RJ part of the spectrum without any information on the Wien side. The spectral index is constrained by the slope of the RJ portion of the grey body while the temperature is constrained by the peak location. 
In our simulation, we fix the spectral index to 2. The 70$\mu$m band is usually not included in ISM studies because is very sensitive to very small grains (VSG) emission. We include it in this simulation to show the dependence of the parameter recovery on the error bars and on the SED sampling. }

To fit the SED of the simulated sources and to study the parameter distribution in the parameter space, as well as their correlation, we use a Monte Carlo Markov Chain algorithm~\cite[MCMC, ][]{Lewis02}.
Hereafter, we explain the main idea behind this approach. 

Following Bayesian statistics, the probability to have a set of parameters \textbf{p} given a set of data \textbf{d} is

\be\label{eq:mcmc}
P(\textbf{p}|\textbf{d}) \propto P(\textbf{p})P(\textbf{d}|\textbf{p}).
\ee

\noindent Here $P(\cdot | \cdot)$ are conditional probability densities with argument on the right-hand side of the bar assumed to be true, $P(\textbf{p})$ is the {\em a priori} probability density, or prior, of the parameters set \textbf{p},
$P(\textbf{d}|\textbf{p})$ (e.g. the likelihood function) is the probability density of dataset \textbf{d} given a set of parameters \textbf{p} and is described by the formula 

\be
P(\textbf{d}|\textbf{p}) \propto \exp\left({-\frac 12 \sum_{b=1}^{N_{\lambda}}}\left(\frac{d_b - I_b({\textbf p})}{\sigma_{b}}\right)^2\right)
\ee 

{ \noindent where $N_{\lambda}$ is the number of bands, $I_b$ is the value of the chosen model with parameters {\bf p} in the $\lambda_{b}$ band, either Eq.~(\ref{eq:Inu}) or Eq.(~\ref{eq:Inu2}) and $\sigma_{\lambda}$ is the statistical error associated to the 
flux ${d_{b}}$. If Eq.~(\ref{eq:Inu}) is chosen, then $\mathbf p =[\epsilon_0, T_d, \beta]$; otherwise, if Eq.~(\ref{eq:Inu2}) is chosen, then $\mathbf p =[\epsilon_1, \epsilon_2, T_1, T_2]$.
%The {\em a priori} probability densities take into account the joint distribution of the parameters, including their correlations as discussed later.}
{ The {\em a priori} probability density of a given set of parameters is the probability distribution that expresses the parameter uncertainties before the data are taken into account. When nothing is known about the parameter distributions, usually wide flat priors are adopted, as they assign equal probabilities to all possibilities. If, on the contrary, the parameter statistical distribution is known from previous experience, i.e. previous measurements, then a more specific, informative prior might be adopted.    }
%definite information about a variable 

%takes into account the expected distribution of the parameters, as well as their correlations.}

The posterior probability $P(\textbf{p}|\textbf{d})$ is estimated using the MCMC algorithm. 
Given a set $\mathbf{p}_i$, with likelihood $L_i$ and posterior 
probability $P( \textbf p_{i}  | \textbf{d} )$, the MCMC algorithm peforms a random walk through the allowd parameter space, 
generating a new independent set $\mathbf{p}_{i+1}$
with likelihood $L_{i+1}$ and posterior 
probability $P( \textbf p_{i+1}  | \textbf{d} )$. 
This second set is 
accepted according to a rule which also guarantees a good sampling 
of the probability density in a reasonable computational 
time. 
We make use of the
Metropolis-Hastings algorithm, which is a class of Markov Chains in which the new set $\mathbf{p}_{i+1}$ 
is always accepted if 
%============================
\be\label{eq:mcmc1}
\Lambda(i+1,i) = \frac{P( \textbf p_{i+1}  | \textbf{d} ) }{ P ( \textbf p_i | \textbf{d} ) } =
\frac{L_{i+1}P(\textbf{p}_{i+1})}{L_{i}P(\textbf{p}_i)} > 1
\ee
%============================
This guarantees the convergence 
to the maximum of the likelihood function. If instead $\Lambda(i+1,i)<1$, the new
set is accepted with a probability proportional to the ratio $\Lambda(i+1,i)$. 
An advantage of this method is to ensure a good sampling of the parameter distribution in parameter space. 
{For an application of this technique on millimeter and sub-millimeter observations, see~\cite{Veneziani10}. }

Fig.~(\ref{fig:punti_nocorr}) shows the 68$\%$ and 95$\%$ posterior probabilities in the T$_d-\beta$ parameter space obtained for the two cases of ISM (T$_{ISM} = 15$K, $\beta=2$) and HII region (T$_{HII}=80$K, $\beta=2$), respectively. 
The elongated and slant shape of these posterior probabilities indicate that the two parameters are degenerate in parameter space.
In case of warm sources we have a strong constraint on $\beta$ but a not precise determination of the temperature, as shown in the right panel. On the contrary, in the case of ISM sources (left panel), the whole SED is well sampled, leading to a smaller degeneracy between the temperature and the spectral index.
%In Fig.~(\ref{fig:cont_nocorr}) we show the posterior probabilities for sources with 5$\%$ (blue line), 10$\%$ (grey line) and 20$\%$ (red line) error bars. It is clear that a spurious anticorrelation arises when the flux noise increases.    

\begin{figure*}[!t]
\begin{center}
\includegraphics[width=7cm]{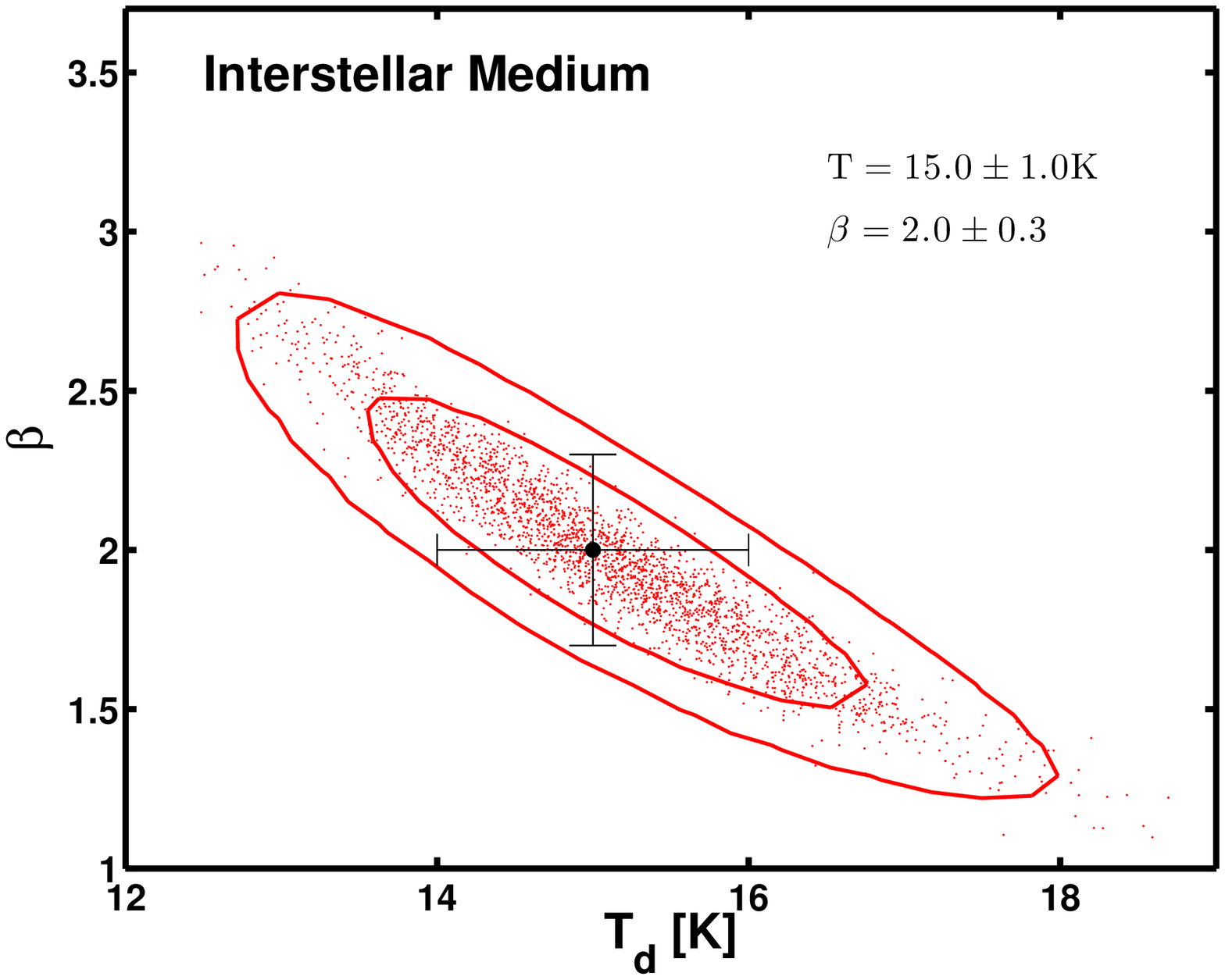}
\includegraphics[width=7cm]{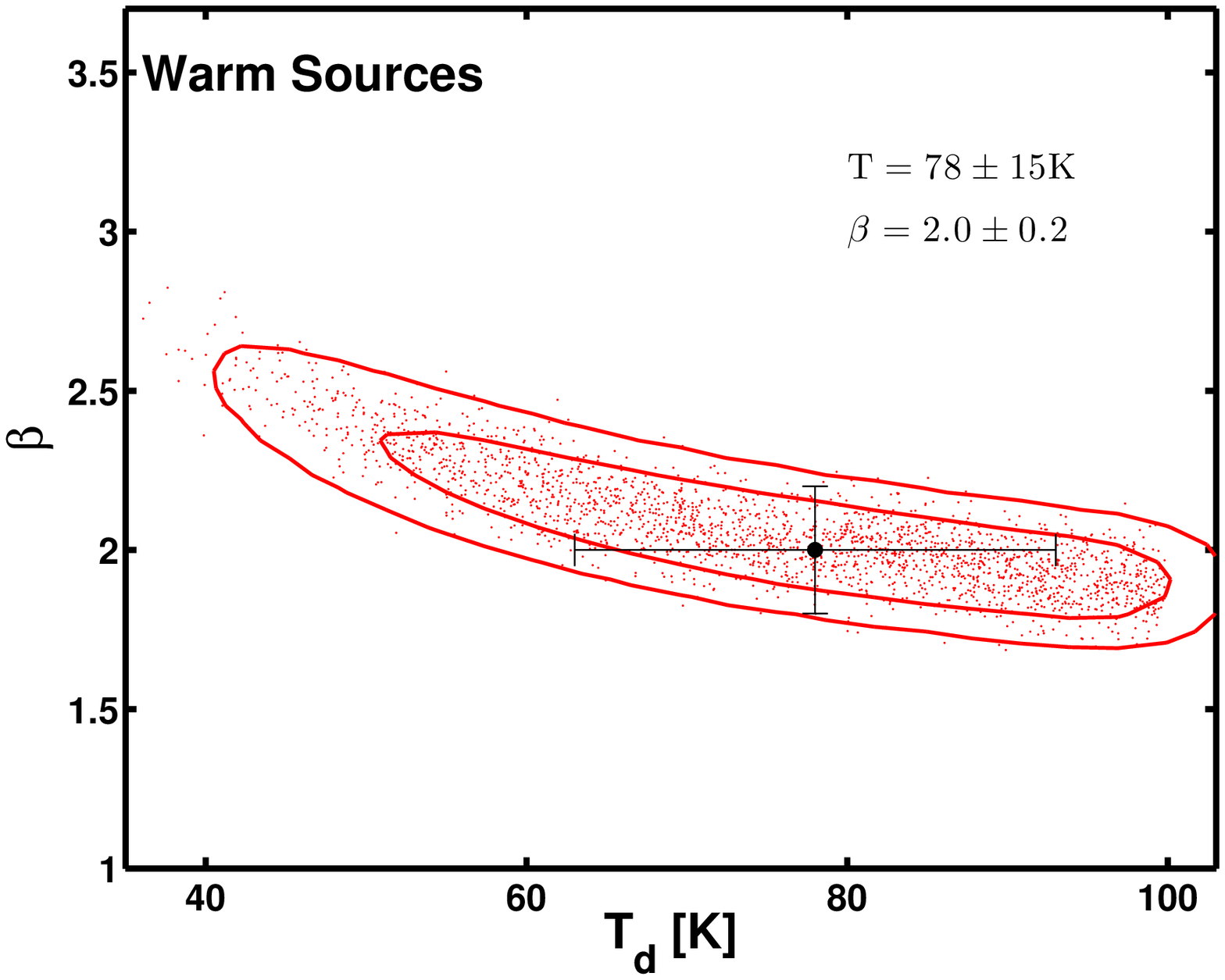}
\caption{68$\%$ and 95$\%$ contours of the two-dimensional posterior probabilities of sources at T = 15 K, $\beta = 2$ (left panel) and T = 80 K, $\beta =2$ with a 20$\%$ statistical error on fluxes. The independent points of the Markov Chains are denser within the 68$\%$ contour, close to the maximum of the distribution, than close to the boundaries. The slant shape means that T and $\beta$ are degenerate and the elongation is larger for the warm sources where Herschel bands do not constrain the peak of the SEDs. The 1$\sigma$ error bars are obtained by marginalizing the tri-dimensional posterior probability, as we have three parameters, over the remaining two dimensions. } 
\vspace{0.5cm}
\label{fig:punti_nocorr}
\end{center}
\end{figure*}

The dependence of one of the two parameters on the other,
i.e. their statistical degeneracy, can be easily demonstrated. Starting
from a one-temperature modified black body:

\be
F(T_b, \beta) = \epsilon_0 \left( \frac{\lambda}{\lambda_0} \right)^{-\beta}
\frac{2 h c^2}{\lambda^5} \frac{1}{e^{\frac{hc}{\lambda k T_d}}-1}\Delta\Omega
\ee
In the R-J regime, we can approximate with:
\be
F(T_b, \beta) \simeq \epsilon_0 \left( \frac{\lambda}{\lambda_0} \right)^{-\beta}
\frac{2 c k T_d}{\lambda^4}
\Delta\Omega
\ee
which is a linear relation in log$\lambda$-log$F$ scale. Thus we expect some
level of anticorrelation between T$_d$ and $\beta$, as in any linear relation there
is some level of anticorrelation between slope and intercept.

The more the fluxes are noisy and the SED not well sampled, the more the two parameters are correlated and difficult to constrain. Moreover, when combining
data of different sources to recover correlation between parameters, there is the tendency to treat systematic errors as if they were statistical, which is not correct. 
Here we treat the systematic errors in a proper way, and use Bayesian statystics, in order to provide a correct reconstruction of the parameters distributions and of physical 
correlations between parameters.

\subsection{Bayesian treatment of systematic errors}\label{sec:syserr}

The sources simulated in Sec.~(\ref{sec:noise}) have up to $20\%$ of statistical uncertainty on measured fluxes. 
Nonetheless, the flux uncertainty might come also from the calibration error and other sources of systematic error, which do not have the same properties as the statistical errors (i.e. instrumental noise, background fluctuations). %The statistical error, coming from the instrumental noise, associated to fluxes is then small and the noise induced T-$\beta$ anticorrelation in one-temperatures sources is negligible. %In this cases, the previous method is redundant and time consuming and the SEDs can be fitted with a normal least square procedure. 
In order to properly take into account the systematic effects, we make use of a Monte Carlo procedure. In the following we will refer only to the systematic error due to calibration uncertainty but the same procedure can be applied to any source of systematic error. 

We would like to emphasize that this treatment is particularly important in our specific case: the recovery of a relationship among physical parameters of 
different observables, several sources in our case. The systematic error can be the same in all sources, as in the case of a wrong calibration of one band. If this is 
not considered, the result can be severely biased. 

In order to properly take into account in our analysis the calibration error 
we use a {\em conditional inference} technique. 
We consider a n-tuple of the calibration values $\mathbf k = (k_1, k_2, ..., k_{N_{\lambda}})$. Each value $k_b$ has
a probability distribution $P(k_b)$, which we assume to be uniform in the $-\Delta k_b<k_b< \Delta k_b$ range. The values of $\Delta \mathbf k$ are reported in Tab.~(\ref{tab:exp}) for each band. 
Given the observed dataset $\mathbf d$, we have a joint distribution of the parameters $\mathbf p$ for all possible configurations of $\mathbf k$, 
$P(\mathbf p | \mathbf d , \mathbf k)$.
The conditional result {for a single source} is obtained by marginalization of this probability over all the possible values of the calibration uncertainties, weighted by 
their distribution:

{
\begin{equation}\label{eq:cond_prob}
P(\mathbf p | \mathbf d) = \int_{-\Delta k}^{\Delta k} P(\mathbf d, \mathbf k | \mathbf p) P(\mathbf p) P(\mathbf k) d\mathbf k
\end{equation}
\noindent where
\be\label{eq:like}
P(\textbf{d, k}|\textbf{p}) \propto \exp\left({-\frac 12 \sum_{b=1}^{N_{\lambda}}\left(\frac{d_b -  k_b I_b({\mathbf p})}{\sigma_b}\right)^2}\right)
\ee

}

Eq.~(\ref{eq:cond_prob}) is an exstention of Eq.~(\ref{eq:mcmc}) where both statistical and systematic uncertainties are taken into account.  
In order to apply this technique, we run a Monte Carlo simulation, scattering the values of $\mathbf k$ in 100 realizations. We tried different numbers of calibration realizations, but we could not find any difference in the final results above 100 steps.
In each iteration $\mu$, a calibration uncertainty vector $\mathbf k_\mu$ is generated. 
The fluxes {\bf d} of all sources are multiplied by the same set of $\mathbf {k_\mu}$. The fit of the SED in each iteration is performed with a Monte Carlo Markov Chain algorithm (see Sec.~\ref{sec:noise}). %The assumed {\em a priori} probability densities of the parameters are the same for all sources and for every realization of $\mathbf k$, and are reported in Tab.~(\ref{tab:priors}). 

The final physical parameters which better describe the SED of each source are then obtained by marginalizing the 100 calibration dependent set of physical parameters over the calibration uncertainties as in Eq.~(\ref{eq:cond_prob}).  
The marginalization is performed with the publicly available GetDist software~\citep{Lewis02}.

\subsection{Method description and uncertainties analysis}%\label{sec:method}

{ 
The method to recover dust physical parameters focuses on our treatment of statistical and systematic uncertainties and on a good definition of the {\em a priori} probability density in the MCMC. 
%We also analyze their effect in inducing spurious anticorrelations among the model parameters. 
%\subsection{Procedure}
The analysis is carried out according to the following steps. The details of each step are reported later in this section.

\begin{enumerate}
\item We make a first, fast MCMC run on the whole sample of sources to check the best model for each SED and to study the parameter distributon in the parameters space. Therefore, we fit the SED of each source with both M$_1$ and M$_2$, taking into account both statistical and systematic uncertainties (i.e. calibration errors) as described in Sec.~(\ref{sec:syserr}). At this stage, we assume uniform wide priors, reported in Tab.~(\ref{tab:priors}). 

\item We determine if M$_1$ or M$_2$ better fit the SED of a source, 
based on the comparison of the $\chi^2$ probability density functions, as reported in Sec.~(\ref{sec:model});
if both models give good probability, we assign the source the simpler M$_1$;
if both models give bad probability we exclude the source from the analysis. An example of good and bad fits after step (1) is reported in Fig.~(\ref{fig:hist_nocorr}).
\item For the sources identified with M$_1$, %and for which we recover physically meaningful values of the parameters, 
we estimate again the physical parameters taking into account both statistical and systematic uncertainties, as in step 1, but this time with a more precise definition of the {\em "a priori"}
probability densities. Therefore, we assume multivariate Gaussian priors, estimated from the first run. This procedure is described in Sec.~(\ref{sec:parest}).
\item From the new physical values the temperature-spectral index relationship is measured as reported in Sec.~(\ref{sec:betat}).  
\end{enumerate}

}

\subsection{Model identification}\label{sec:model}

In real observations, especially of the Galactic Plane, there is often an overlap of sources along the line of sight or a multi-component temperature source. % We are then not fully sure about how many components we are observing. 
It is crucial to be able to distinguish among different models, and identify the correct one, before attempting any physical interpretation. 

As a first step in our analysis, we then fit the simulated dataset with both models in Eq.~(\ref{eq:Inu}) and~(\ref{eq:Inu2}), i.e. with a single temperature and two-temperatures. 
In M$_1$, the fitted parameters are the emissivity ($\epsilon_0$), the temperature T$_d$ and the spectral index $\beta$. 
We do not consider any model with more than two components because with the datasets available nowadays it is unlikely to have enough data points to well fit up to 6 parameters. 
For the same reason, in the two components case, we cannot perform a fit over all the six parameters: T$_1$, T$_2, \epsilon_1, \epsilon_2, \beta_1, \beta_2$.
We rather set the spectral indices to a fixed common value $\beta_c = \beta_1 = \beta_2$, and estimate the two emissivities and the two temperatures. The values explored for $\beta_c$ in model 2 are 1.7, 2.0, 2.3 and 2.7, consistently with~\cite{fds}. We then execute the fit for M$_2$ four times, changing every time the spectral index which is assumed to be the same for both the components, i.e. not temperature dependent. 
{To make sure not to constrain the results, we assume uniform wide {\em a priori} probability densities on every parameter. They are the same for all sources and are reported in Tab.~\ref{tab:priors}}.

\begin{table}[!t]
\begin{center}
\space
\caption{{Priors during the model identification (step (1))}}
\label{tab:priors}
\begin{tabular}{l  c c c}
 & {\bf p} & Range of variability & P({\bf p})\\
\hhline{~===}\\
\multirow{3}{*}{Model 1} & $\epsilon_0$ & $>$ 0 & U\\
&$\beta$ & 0.5-4 & U\\
&T$_d$ (K)& 5-100 & U\\
\hline
\multirow{4}{*}{Model 2} & $\epsilon_1$ & $>$ 0  & U\\
&$\epsilon_2$ & $>$ 0 & U\\
&T$_1$ (K)& 5-25 & U\\
&T$_2$ (K)& 10-100 & U\\
& $\beta_C$ & 1.7, 2.0, 2.3, 2.7 & U\\
\hline\\
\end{tabular}
\end{center}
\footnotesize{{List of {\em a priori} probability densities imposed on parameter set during the model identification procedure.      
We chose wide uniform (U) ranges at values in order not to constrain the fit results and to avoid the code to converge to non physical values. The explored range of $\beta_C$ values is also reported. }}
\vspace{0.5cm}
\end{table}

We can then easily find the model which better fits the SED by using the $\chi^2$ probability density function. Since M$_1$ and M$_2$ have a different number of degrees of freedom, we compare the two analyses using the cumulative distribution function $P$ of a $\chi^2$ distribution with the proper number of degrees of freedom. The ideal fit should have $P\sim0.5$. $P\ll 0.5$ indicates an overestimate of the error bars while $P\gg 0.5$ indicates that the fitting model is not a good model for the dataset. In one-component sources, we expect, in an ideal situation, $P(M_1)$ to be near to 0.5, and $P(M_2)$ to be larger. 
The opposite is true for two-component sources. In an ideal situation, we expect to have at least one of the $P(M_2;\beta_c)$ near 0.5, and $P(M_1)$ to be larger. 

{ 
An example of one and two-dimensional posterior probabilities of the physical parameters of an ISM-like source and of an HII region combined with dust is shown in Fig.~(\ref{fig:hist_nocorr}). In the top line of this figure we show the posterior probabilities of a 20K source (ISM) fitted with a one temperature model (top left panel) and with a two-temperature model (top right panel). The convergence is reached with the one-temperature model because all the three parameters (temperature, spectral index and emissivity) are well constrained within the boundary provided. On the contrary, the fit with two components (top right) is not well constrained. The four parameters (two temperatures and two emissivities) do not converge within the acceptable ranges. In the bottom line of Fig.~(\ref{fig:hist_nocorr}) the fit of a two component sources with a one and two-components model is shown. The source is an HII region of 50K combined with some surrounding dust at 17K. The spectral indices of both components are set to 2. In this case, the fit with a two-components model converges (bottom right) with the four parameters well defined within the chosen ranges, while the fit with a one-component model (bottom left) is poorly constrained within the physically acceptable boundaries.
}

\begin{figure}[!t]
\begin{center}
\setlength\fboxsep{5pt}
\setlength\fboxrule{0.5pt}
\fbox{\includegraphics[width=0.21\textwidth]{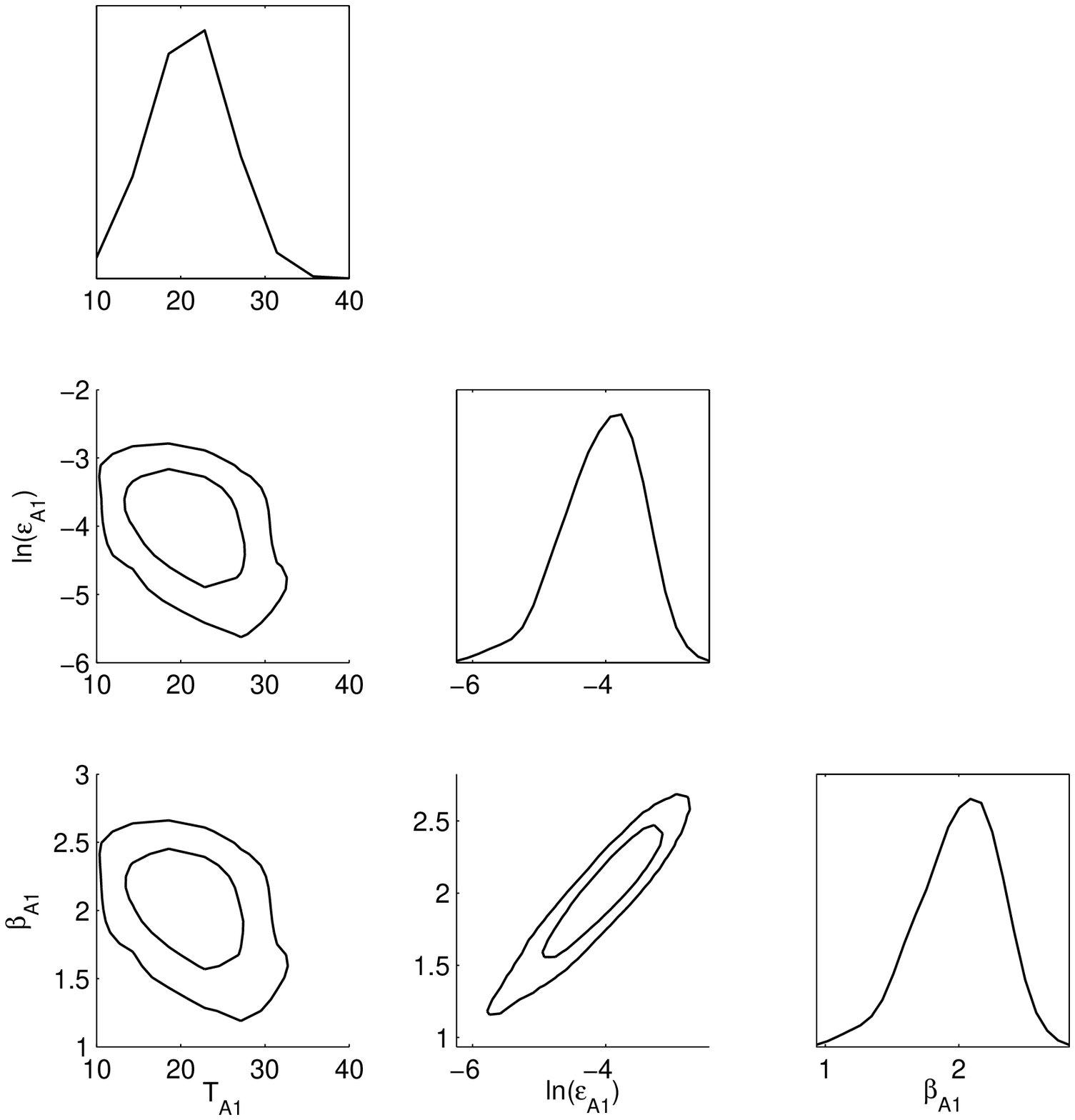}}
\fbox{\includegraphics[width=0.21\textwidth]{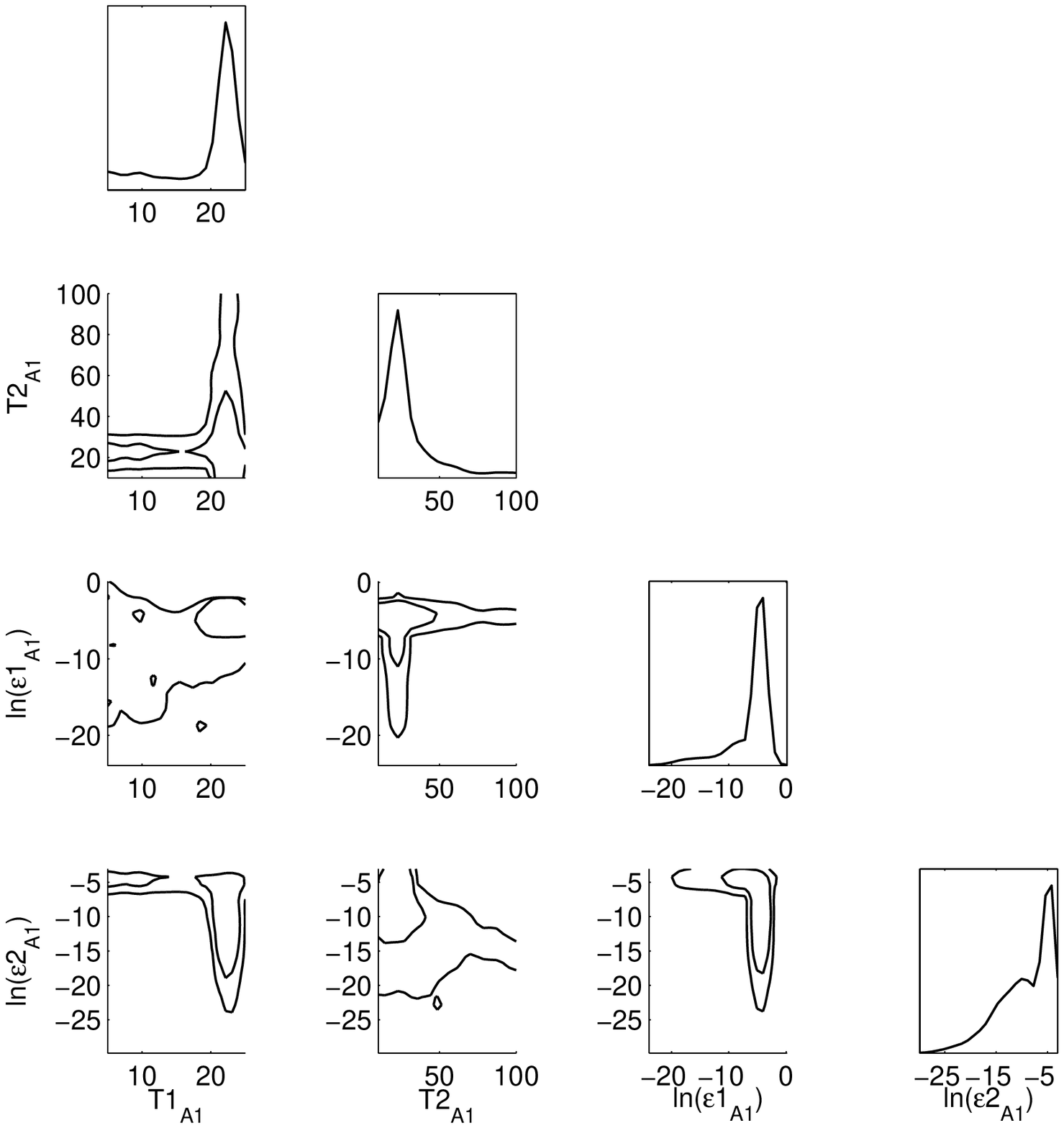}}
\fbox{\includegraphics[width=0.21\textwidth]{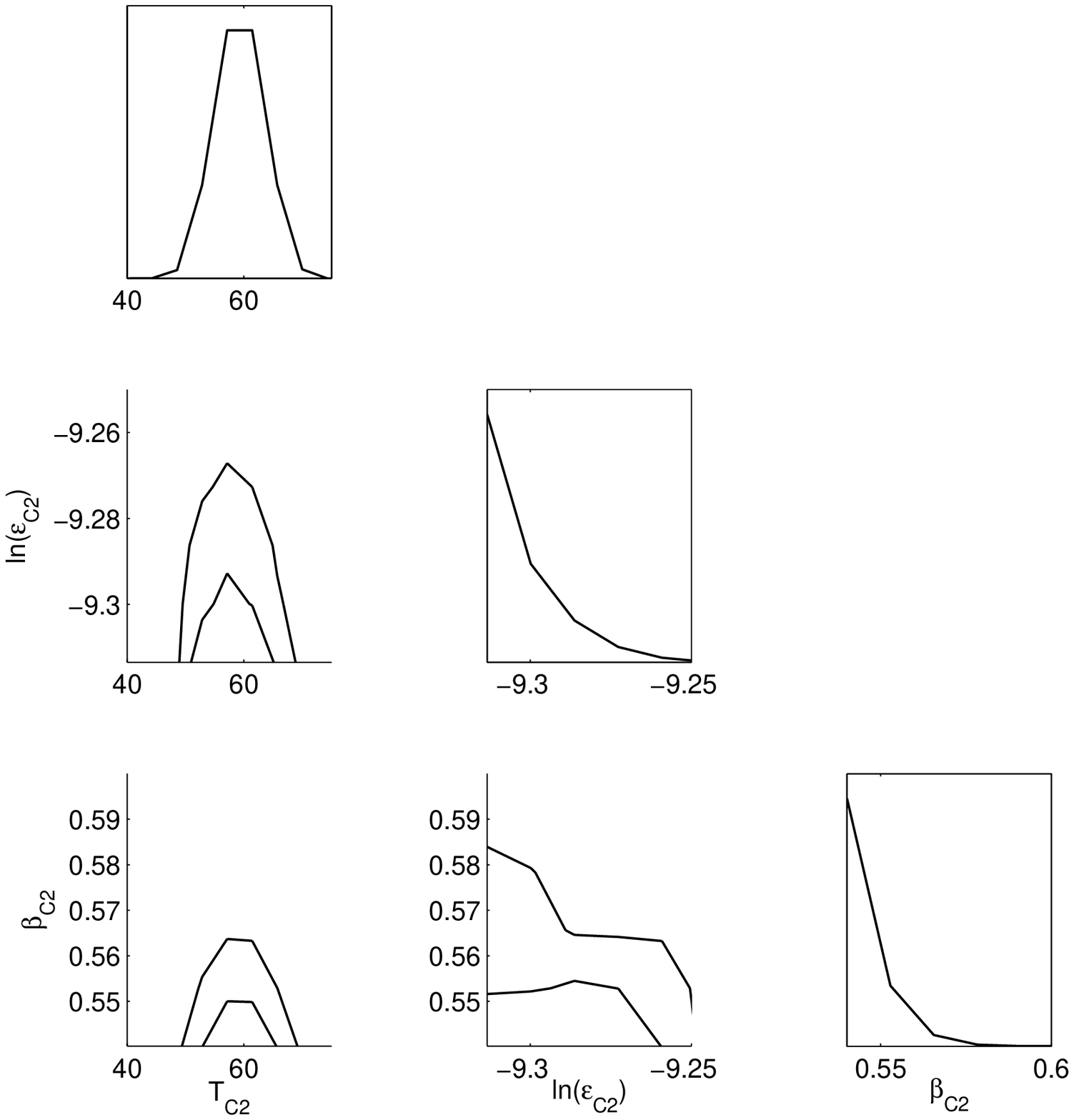}}
\fbox{\includegraphics[width=0.21\textwidth]{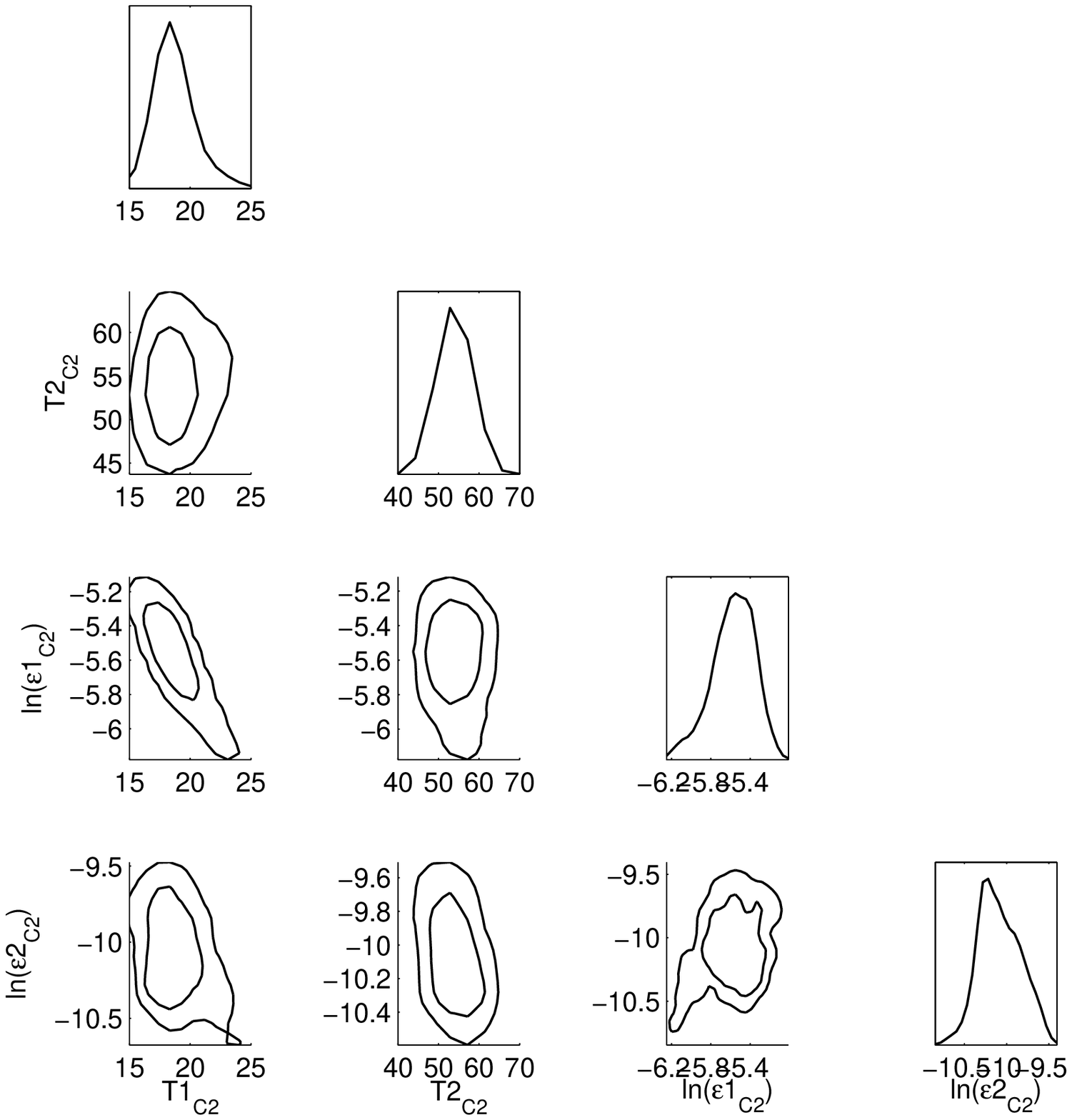}}
\caption{One and two-dimensional physical parameters posterior probabilities for an ISM-like (source A1, see Tab.~(\ref{tab:exp})) and an HII (source C2) region-like source estimated with a MCMC run assuming wide uniform priors. The ISM is well fitted by M$_1$ (top left) and not by M$_2$ (top right). As a consequence the posterior probabilities are well constrained within the range provided when considering a one-temperature model (top left) and not with a two-temperature model (top right). The opposite is true for the second one: M$_1$ is not well constrained (bottom left) and converges to very low value of the spectral index $\beta_{HII}$, while all the unknowns in M$_2$ converge to finite results (bottom right).}
\vspace{0.5cm} 
\label{fig:hist_nocorr}
\end{center}
\end{figure}

{
\subsection{{ Parameter Estimates}}\label{sec:parest}
{ Since the parameters in Eqs.~(\ref{eq:Inu}) and~(\ref{eq:Inu2}) are not uniformly distributed and are also intrinsically correlated in the parameter space, their {\em "a priori"} probability density P({\bf p}) (Eq.~\ref{eq:cond_prob}) must take these properties into account. This is required to avoid biased estimates of the source characteristics. The intrinsic parameter correlation in each source is due to the spectral shape describing the source emission (Eqs.~\ref{eq:Inu} and~\ref{eq:Inu2}) and it doesn't have anything to do with the physical dependency of one parameter on the other, which can be measured from an ensemble of objects.
The level of the intrinsic correlation, in our case, depends on the random noise present in the single SEDs, on the SED sampling within the considered bands and on possible source overlap along the line of sight. Since those effects might vary source by source, we do not perform a fit of the parameters all together, but analyze the parameter statistical distribution of each source independently of the others. The information about the parameter statistical distribution and intrinsic correlation is contained in the covariance matrix, estimated for each source by means of the first MCMC run (step (1)). The covariance matrix is then used to build the new {\em a priori} probability distribution, which is not flat anymore as in the first run, but contains specific and definite information about the variables.     }
{The new set of priors P({\bf p}) is a multi-variate Gaussian, with the center {$\overline{\mathbf p}=[\log{\epsilon_0}, \beta_0, T_0]$} in the approximate average values of the whole sample of sources. The covariances of the multi-variate Gaussian priors are estimated, for each source, by broadening 10 times the parameters standard deviations keeping their correlations. This makes us sure not to constrain the final results with a too tight parameters range. 
In other words, while the center of the Gaussian prior are the same for the whole set of sources, the parameter covariances are estimated source by source and, as a consequence, the fit is performed source by source.   
An example of multi-variate Gaussian priors for a single ISM-like, cold clump-like and HII region-like source, is reported in Tab.~(\ref{tab:priors2}) and shown in Fig.~(\ref{fig:priors2}). }

Taking all those aspects into account, the final parameters estimate on each source is then performed again by applying Eq.~(\ref{eq:cond_prob}) and~(\ref{eq:like}) but, this time, instead of assuming uniform distributed priors, we describe P({\bf p}) as a multi-variate Gaussian:

\be\label{eq:priors}
P(\mathbf p) \propto \exp\left({-\frac 12 \left((\mathbf p-\overline{\mathbf p})^T\Sigma^{-1}(\mathbf p-\overline{\mathbf p})\right)}\right)
\ee
\noindent where 
$\Sigma_{ij}$ % = \left<(d-I_\lambda)_i(d-I_\lambda)_j\right>$ 
is the covariance matrix of each source, with the indices $i$ and $j$ running over the entire set of  paramers. $\Sigma_{ij}$ elements are obtained from the correlation matrix $C_{ij}$ by $\Sigma_{ij}= C_{ij}\;\sigma_i\;\sigma_j$ where $\sigma_i\;\sigma_j$ are the standard deviations of the parameters $p_{i}$ and $p_{j}$, obtained from the first MCMC run, multiplied by a factor of 10 in order to broaden the prior, retaining the correlation properties.

An example of $C$ elements are reported in Tab.~(\ref{tab:priors2}). $\Sigma$ contains therefore the information about the distribution and correlation of parameters. Both $\overline{\mathbf p}$ and $\Sigma$ are estimated through the MCMC in step (1). The final likelihood is then

\bea\label{eq:flike}
%P(\mathbf p | \mathbf d) \propto  \int_{k} \exp (-\frac{1}{2}  \sum_{b = 1}^{N_{\lambda}}\left(\frac{d_\lambda -  k_\lambda I_\lambda({\mathbf p})}{\sigma_{\lambda}}\right))
P(\mathbf p | \mathbf d) \propto  & \displaystyle \int_{-\Delta \mathbf k}^{\Delta \mathbf k} \exp \left({-\frac{1}{2} \sum_{b = 1}^{N_{\lambda}}\left(\frac{d -  k I({\mathbf p})}{\sigma_{d}}\right)^2_{b}}\right)\cdot \nonumber  \\
&\displaystyle \exp\left({-\frac 12 \sum_{i=1}^{N_{p}}\sum_{j=1}^{N_{p}}\left(\frac{(p_i-\overline{p_i})C^{-1}(p_j-\overline{p_j})}{\sigma_{i}\sigma_{j}}\right)}\right) d\mathbf k
\eea

\noindent where $N_{p}$ is the number of parameters and $\mathbf k$ is the vector of the band calibration uncertainties on extended emission. 
As already pointed out, with our method sources are fitted one by one in order to take into account different values of the covariance matrix, due to i.e. different random noise or different sampling of the SEDs, and different systematics effects on different sources, i.e. overlap of other objects on the line of sight.  

{ If the sources in the sample have similar noise level and systematic characteristics, i.e. line-of-sight temperature variations and calibration uncertainties, one can also use a hierarchical Bayesian procedure~\citep{Kelly12}. The hierarchical model fits the SEDs of the whole sample of sources at the same time, assuming a global model for the overall distribution of source parameters. }

}
\subsection{Estimate of the Temperature-Spectral Index relationship}\label{sec:betat}

As discussed in the Introduction, an inverse relation between temperature and spectral index has been found in several previous studies. We model this relation by means of Eq.~(\ref{eq:betat}).

The comparison between M$_1$ and M$_2$ {and the estimate of the functional form of the parameter distributions, are the first steps to estimate the temperature-spectral index relationship.  }
We select as possible source candidates of a temperature dependent spectral index, the ones with M$_1$ preferred over M$_2$ and with $P(M_1) < 0.95$ $(2 \sigma)$.  
This last requirement is necessary because the worst the fit, the more spurious the T$_d-\beta$ anticorrelation. 

If all the previous conditions are satisfied, we estimate the T$_d-\beta$ trend on the selected subsample of sources, taking into account the calibration uncertainties again with a Monte Carlo procedure on the calibration uncertainties. 
In each iteration $\mu$, i.e. for each $\mathbf k_{\mu}$, we fit sources with M$_1$ through Eq.~(\ref{eq:cond_prob}) and Eq.~(\ref{eq:priors}) and then estimate Eq.~(\ref{eq:betat}) {by means of a MCMC run assuming uniform wide priors on $A_{\mu}$ and $\alpha_{\mu}$: 0 < A < 3, $-2<\alpha<2$.} 
%. In each iteration of the Monte Carlo, we fit equation~\ref{eq:betat} over the sources which are consistently scattered with the same set of calibration factors $\mathbf k$. 
After 100 iterations, we have 100 pairs of $A_{\mu}$ and $\alpha_{\mu}$. 
We then marginalize over the calibration errors as in Eq.~(\ref{eq:cond_prob}):
\begin{equation}
P(A, \alpha | \mathbf x) = \int_{-\Delta \mathbf k}^{\Delta \mathbf k} P(\mathbf x; \mathbf k | A, \alpha) P(A, \alpha)P(\mathbf k) d \mathbf k
\label{eq:cond_prob2}
\end{equation}

\noindent where $\mathbf x = (T_d, \beta)$, $P(\mathbf k)$ is the same as in Eq.~(\ref{eq:cond_prob}), $P(A,\alpha)$ is assumed to be uniformly distributed and uncorrelated, and the likelihood of the $\mu$-th iteration is
\be
\begin{split}
%P(A, \alpha | \mathbf T_d, \beta; \mathbf k)\propto \exp\left(-\frac 12(\frac{\beta-A(\frac{T_d}{T_0})^\alpha}{\sigma_\beta})_i\right)
P(\mathbf x; \mathbf k | A, \alpha)\propto 
\exp\left(-\frac 12 \sum_{s=1}^{N_{s}} \left (\frac{\beta(\mathbf k_{\mu})-A_{\mu}\left[\frac{\mathbf T_d(\mathbf k_{\mu})}{T_0}\right]^{\alpha_{\mu}}}{\sigma(\beta(\mathbf k_{\mu}))}\right)_{s}^2\right) 
\end{split}
\ee

\noindent {where $N_{s}$ is the number of sources, $\beta(\mathbf k_{\mu})$ and $T_{d}(\mathbf k_{\mu})$ are the spectral index and temperature of source $s$, estimated with the calibration uncertainties $\mathbf k_{\mu}$ and ${\sigma(\beta_{k})}$ is the statistical error on $\beta(\mathbf k)$.}
One dimensional probabilities of A and $\alpha$ are obtained by marginalization over the other parameter.

\section{Simulated observations}\label{sec:sources}

Since we are interested in studying the SED fitting using the Herschel bands, as a baseline of our analysis we use the PACS and SPIRE wavelengths, angular resolutions and noise properties. 
We complement the Herschel spectral coverage simulating fluxes also in the MIPS (24$\mu$m), IRAS (100$\mu$m) and Planck-HFI (850$\mu$m) bands. These bands are particularly helpful to constrain the SEDs of warm and cold single and multi-temperatures objects whose emission is not fully sampled by the PACS and SPIRE instruments.
{Depending on the model, we simulate one or two-temperatures modified black body with a 1-$\sigma$ statistical uncertainty of $\sigma_{\lambda}=10\%$ of the total flux in each band. This value arises from the combination of the experimental Herschel noise confusion limit (few percent of the total flux) combined with some non-negligible background fluctuation and it's the value also chosen by~\cite{Shetty09a} in their analysis. }The calibration uncertainties are estimated on diffuse emission. 
The fluxes are then scattered within the error bars twice: first, in a Gaussian uncorrelated fashion, to take into account random noise; second, in a correlated fashion, i.e. the same bands have the same scatter in all sources, to take into account the calibration uncertainties.
We assume a common angular resolution, i.e. 4', for IRAS 100$\mu$m, or 5' when Planck 850$\mu$m is also included. 
The considered bands and the experimental characteristics used to perform the simulations are reported in Tab.~(\ref{tab:exp}).

\begin{table*}[!t]
\begin{center}
\space
\caption{Simulated observing experiments}
\label{tab:exp}
\begin{tabular}{c c c c c c}
Band ($\mu$m) & Experiment & Beam size & $\Delta k_{ext}$ &  $\Delta k_{point}$ &$\sigma$\\
\hhline{~=====}\\
24 &  MIPS~\citep{Carey09} & $6" $ & {15$\%$} & {4$\%$} &{10$\%$}\\
70 & PACS~\citep{Poglitsch10} & 6" & {20$\%$} & {3$\%$} &{10$\%$}\\
100 & IRAS-IRIS~\citep{MamD05}& 4' & {13.5$\%$} & {13.5$\%$}& {10$\%$}\\
160 & PACS~\citep{Poglitsch10} & 11" & {20$\%$} & {4$\%$}& {10$\%$}\\
250 & SPIRE~\citep{Griffin10} & 18" & {15$\%$}& {7$\%$} & {10$\%$}\\
350 & SPIRE~\citep{Griffin10} & 25" & {15$\%$} & {7$\%$}& {10$\%$}\\
500 & SPIRE~\citep{Griffin10} & 37" & {15$\%$} & {7$\%$}& {10$\%$}\\
850 & Planck-HFI~\citep{planck_mission} & 5' & {3$\%$}& {3$\%$}& {10$\%$}\\
\hline
\end{tabular}
\end{center}
\footnotesize{Experimental characteristics used to generate the simulated sources. {The reported calibration uncertainties ($\Delta k_{ext}$ and $\Delta k_{point}$) refer to diffuse emission and to point sources, respectively. }The statistical error $\sigma$ is assumed to be 10$\%$, which is the maximum value obtained in the analysis of Hi-Gal data. It includes both instrumental noise and background fluctuations. } 
\vspace{0.5cm}
\end{table*}

We simulate observations of sources with different physical conditions. 
The simplest case consists of one-component sources, described by Eq.~(\ref{eq:Inu}), with a temperature around 20K and a spectral index which may or may not vary with temperature (cases A$_1$ and A$_2$, respectively). In the following, we will refer to the one-component model as model 1 (M$_1$). A more complex case is given by the overlap of more than one source along the line of sight. 
Indeed, one of the key questions of the SED fitting is the error made when approximating a multi-temperature observation with an isothermal model and if this approximation can lead to a spurious correlation among the fitted parameters~\citep{Shetty09b}. This could be the case when analyzing a source with more than one population of dust grains or more than one source along the line of sight. We have considered two different scenarios, one in which the two temperatures are very close (few degrees apart) and one in which they are very different (tens of degrees). A typical source of the first case is a cold dense pre-stellar clump, i.e. a cold clump with a temperature (T$_C$) around 10K surrounded by an envelope of dust with an average temperature (T$_E$) of around 15K (case B). A typical source of the second case might be an HII region combined with foreground/background diffuse dust (case C). Both the considered scenarios are described by Eq.~(\ref{eq:Inu2}) and, in the following, we will refer to the two-component model as model 2 (M$_2$).

These last two cases are also chosen to investigate the effect of the band coverage. % on the model identification.
A good sampling of the SED peak and of the Rayleigh-Jeans (RJ) side is required, in order to properly fit both temperature and spectral index and to identify the underlying physical model.  %temperature-spectral index analysis. For cold sources, 
For example, in HII regions, the Herschel wavelength coverage is not enough to separate the two components, while including the MIPS 24$\mu$m the two components are properly detected. Similarly, pre-stellar clumps properties are better identified including Planck-HFI 850$\mu$m band.

The simulated sources are described below and more details are provided in Tab.~(\ref{tab:sources}):
%We simulate three different kind of sources, summarized in table~\ref{tab:sources}: 

\begin{itemize}
\item[{$ \left[\mathrm A \right]$}] ISM environment, as in model 1: single component sources with a Gaussian temperature distribution centered in 20K and ranging between 15K and 25K, observed with a 4' resolution in the spectral range 100-500$\mu$m. We place ourselves exactly in the same situation as~\cite{Paradis10} in order to test our ability to recover the real relationship between temperature and spectral index. 
{For this purpose, we first set the spectral index to 2 (sources A$_1$) and then we use as input a temperature dependent spectral index, both anticorrelated, as in~\cite{Paradis10}, (sources A$_2$) and correlated positively correlated (sources A$_3$).}
 
\item[{$ \left[\mathrm B \right]$}] Cold clumps (CC), based on model 2: two-component sources consisting of an envelope at a temperature T$_E$ and a core at a temperature T$_C$. The spectral indices of both components are set to 2. We investigate two options for the relative emissivities as analyzed by~\cite{Shetty09b}. We also study the dependence of the fit on the spectral coverage by measuring the emission in the range 100-500 $\mu$m and then including also the 850$\mu$m flux which constrains the RJ side of the SEDs. % and force the observation to be simulated at 5' resolution.
\item[{$ \left[\mathrm C \right]$}] HII regions combined with foreground/background diffuse dust, of temperatures T$_{HII}$ and T$_{dd}$, respectively. This case is also based on model 2. The spectral indices are set to 2 and the relative emissivities are scaled according to the analysis performed on Hi-GAL data by \cite{Paladini12}. As in the case of the cold clumps, we investigate the dependence of the physical parameters recovery on the spectral coverage. We then sample the SEDs in the 70-500$\mu$m range and then include also the 24$\mu$m flux to better constrain the temperature of the warmer component. In this analysis we assume the 24$\mu$m emission to be dominated by Big Grains (BG) as recently found, for example, by SOFIA~\citep{salgado2012}. The BG thermalize with the Interstellar Radiation Field (ISR) and we can model their emission as a modified grey body. If, on the contrary, the 24$\mu$m flux is dominated by VSG, we need to remove this point from the analysis and consider only the $70-500\mu$m range, because they don't have the same physical properties as BG.  
\end{itemize}

In order to compare the temperatures obtained fitting with M$_1$ the SEDs of two-temperature sources with the input values, we make use of a density-weighted temperature introduced by~\cite{Doty02}, $T_{col}$. According to their definition, the column temperature of a source described by Eq.~(\ref{eq:Inu2}) is given by:
\be\label{eq:tcol}
T_{col} = \frac{\epsilon_1 T_1+\epsilon_2 T_2}{\epsilon_1+\epsilon_2}
\ee

\begin{table*}[!t]
\begin{center}
\space
\caption{Simulated sources}
\label{tab:sources}
\begin{tabular}{l c c c c c }
\hline
Input & Central temperature & Relative column & $\beta$ & Resolution & Spectral Range \\ % & {T-$\beta$} \\
sources & (K) & densities &   &  & ($\mu$m) \\
\hhline{~=====}\\
Interstellar Medium (A$_1$) & T$_A$= 20 & &2 & 4' & 100-500\\ % & No\\
Interstellar Medium (A$_2$) & T$_A$ = 20 & &$2.2(T_A/T_0)^{-1.3}$ & 4' & 100-500\\ % & Yes\\
Interstellar Medium (A$_3$) & T$_A$ = 20 & &$2(T_A/T_0)^{+0.5}$ & 4' & 100-500\\ % & Yes\\
Cold Clump (B$_1$) & T$_C = 10$, T$_E = 15$, T$_{col} = 12.5$ & N$_C$ = N$_E$ & 2 & 5' & 100-500\\ % & No\\ 
Cold Clump (B$_2$) & T$_C = 10$, T$_E = 15$, T$_{col} = 12.5$ & N$_C$ = N$_E$ & 2 & 5' & 100-850\\ % & No\\ 
Cold Clump (B$_3$) & T$_C = 10$, T$_E = 15$, T$_{col} = 10.9$ & N$_C$ = $10$ N$_E$ & 2& 5' & 100-500 \\ % & No\\ 
Cold Clump (B$_4$) & T$_C = 10$, T$_E = 15$, T$_{col} = 10.9$ & N$_C$ = $10$ N$_E$ & 2& 5' & 100-850 \\ % & No\\ 
HII+dust (C$_1$) & T$_{HII} = 17$, T$_{dd} = 50$, T$_{col} = 17.3$ &  N$_{dd}$ = $100$ N$_{HII}$ & 2 & 4' & 70-500\\ % & No\\ 
HII+dust (C$_2$) & T$_{HII} = 17$, T$_{dd} = 50$, T$_{col} = 17.3$ &  N$_{dd}$ = $100$ N$_{HII}$ & 2 & 4' & 24-500\\ % & No\\ 
%HII+dust (D$_2$) & T$_1 = 17$, T$_2 = 50$ &  N$_1$ = $100$ N$_2$ & 2  \\ % & No\\ 
\hline
\end{tabular}
\end{center}
\footnotesize{Three types of simulated sources used to test our method. In each model we generate 1000 sources with a gaussian distribution in temperature. The distribution is contained on the value reported in the 2nd column, with a dispersion of 10$\%$. In case A$_2$ there is a T$_d-\beta$ anticorrelation in input. $T_0$ is set to 20K. } 
\vspace{0.5cm}
 %within the indicated range in cases A and B. In models C and D we perform a gaussian scattering around the reported value within a dispersion of 10$\%$. }
\end{table*}

\section{Results}\label{sec:results}

%In Table~\ref{tab:1comp} we report the analysis of all the sources. In fact, 
In this section we present the results of the SED fitting, and the T$_d-\beta$ relationship 
estimate after properly taking into account the systematic 
errors of the datasets, for each type of source considered.

{The multi-variate Gaussian priors of one sample source for each set are reported in Tab.~(\ref{tab:priors2}) and shown in Fig.~(\ref{fig:priors2}). }

\begin{table}[!t]
\begin{center}
\space
\caption{{Priors during the final estimate (step (3))}}
\label{tab:priors2}
\begin{tabular}{l  c c l l c}
% Sources & Param & Center & \multicolumn{2}{c} {Covariance} & Distribution\\
  Sources & {\bf p} & $\overline{\mathbf p}$ &$\sigma_{\overline{\mathbf p}}$ &\multicolumn{1}{c}{$C$} & P({\bf p})\\
\hhline{~=====}\\
\multirow{2}{*}{ISM} & $\log{\epsilon_0}$ & -4.2 & 9.7 &$C$($\log\epsilon_0$,$\beta$)=0.985  & MVG\\
\multirow{2}{*}{(A)}&$\beta$ & 1.8  &  2.5 &$C$($\log\epsilon_0$,T$_d$)=-0.985 &MVG\\
&T$_d$ (K)& 23 & 85.7 &$C$($\beta$,T$_d$)=-0.956 &MVG\\
\hline
\multirow{2}{*}{CC} & $\log{\epsilon_0}$ & -6.5 & 6.1 &$C$($\log{\epsilon_0}$,$\beta$)=0.982& MVG\\
\multirow{2}{*}{(B)}&$\beta$ & 1.8 & 1.6&$C$($\log{\epsilon_0}$,T$_d$)=-0.986 & MVG \\
&T$_d$ (K)& 13 & 54.8 &$C$($\beta$,T$_d$)=-0.951& MVG\\
\hline
\multirow{2}{*}{HII+ISM} & $\log{\epsilon_0}$ & -5.5 & 34.9 &$C$($\log{\epsilon_0}$,$\beta$)= 0.969& MVG\\
\multirow{2}{*}{(C)}&$\beta$ & 1.8 & 9.0&$C$($\log{\epsilon_0}$,T$_d$)= -0.976& MVG\\
&T$_d$ (K)& 23 & 309.3&$C$($\beta$,T$_d$)= -0.924& MVG\\
\hline\\
\end{tabular}
\end{center}
\footnotesize{{List of {\em a priori} probability densities imposed on the final paremeter estimates. The covariance matrix ($\Sigma$) elements are obtained by multiplying the correlation matrix elements ($C$) with the standard deviation of the two correspondent parameters. The values reported are for just one single source for each kind of simulated observations. The overall shape of the probability density is a multi-variate Gaussian (MVG) where we only take the positive values. }}
\vspace{0.5cm}
\end{table}

\begin{figure}[!t]
\begin{center}
\includegraphics[width=0.5\textwidth]{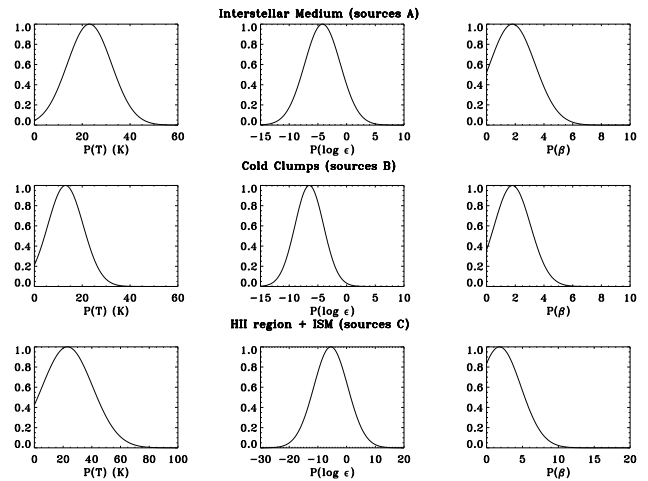}
\caption{One-dimensional projection of the multi-variate Gaussian prior P({\bf p}) described in Eq.~(\ref{eq:priors}) for each set of sources. The parameters are reported in Tab.~(\ref{tab:priors2}).}
\label{fig:priors2}
\end{center}
\end{figure}

The results are summarized in Tab.~(\ref{tab:1comp}) and Tab.~(\ref{tab:result2}).
In Tab.~(\ref{tab:1comp}), first column, we provide the input models; in the second, third and fourth columns,
{the percent fraction of sources assigned M$_1$ and M$_2$ as well as the excluded sources, respectively, highlighting the true positive 
detections;}
columns five to eight report for which $\beta_c$ value the best M$_2$ fit is obtained. 
{In Tab.~(\ref{tab:result2}) we give the input and measured T$_d-\beta$ relationship obtained with our procedure ("Bayesian output") and with a least square method ("Least Square output"), for comparison, for the sources assigned
M$_1$. { The least square method is based on a $\chi^{2}$ minimization to recover the source parameters and their 1$\sigma$ errors. 
Since this method do not perceive any treatment of the systematic uncertainties, besides adding them in quadrature to the statistical errors, it has been tested on simulations where only a 10$\%$ gaussian error bars are present, without including the additional calibration errors.  The least square minimization is used in both stages, the SED fitting and the T-$\beta$ relationship estimate. In this last step, the statistical uncertainties adopted on data points are the ones obtained from the SED fitting step.  }
{In order to claim a physical anticorrelation with the Bayesian procedure, we require a 3$\sigma$ detection of the $\alpha$ parameter. }}

As a general comment we would like to highlight that the two-component model is not well constrained if the spectral coverage is poor. In order to have a clear detection of more than one component we need to have information both on the RJ and the Wien part of the SEDs. If the second component is very faint or the spectral coverage too small to clearly detect the second peak, the marginalized final parameters obtained with M$_2$ are badly constrained. %, with a chi square much higher with respect to M$_1$. 

\begin{table*}[!t]
\begin{center}
\space
\caption{Compared analysis}
\label{tab:1comp}
\begin{tabular}{c | c c c | c c c c}
Input & \multicolumn{1}{c}{$M_1 (\%)$} & { $M_2 (\%)$ } & {Excluded ($\%$)}&{$\beta_c = 1.7 (\%)$} &{$\beta_c = 2.0 (\%)$} &{$\beta_c = 2.3 (\%)$} &{$\beta_c = 2.7 (\%)$}  \\
% \multicolumn{6}{c}{}  & A & $\alpha$&\multicolumn{1}{c}{recovered?} \\
%\multicolumn{2}{c}{} & $\beta_1,\beta_2=1.7$ & $\beta_1,\beta_2=2.0$ &$\beta_1,\beta_2= 2.3 $&\multicolumn{1}{c}{ $\beta_1,\beta_2= 2.7 $}&\\
\hhline{~=======}\\
Interstellar Medium (A$_1$) & {\bf 100} & 0  & 0 & 0 & 0 & 0 & 0 \\ %Yes \\
Interstellar Medium (A$_2$) & {\bf 98} & 2  & 0 & 2 & 0 & 0 & 0 \\%Yes\\
Cold Clump (B$_1$)  & 100 & {\bf 0} & 0 & 0& 0 & 0 & 0 \\%Yes\\
Cold Clump (B$_2$) & 99 & {\bf 1} & 0 & 0 & 0 & 1 & 0  \\%Yes\\
Cold Clump (B$_3$) & 89 & {\bf 11} & 0 & 0 & 1 & 9  & 1 \\%Yes\\
Cold Clump (B$_4$) & 86 & {\bf 14} & 0 & 0 & 0  & 14  & 0 \\%Yes\\
HII+dust (C$_1$)  & 32 & {\bf 68} & 0 & 1 &  48 & 19 & 0  \\%Yes\\
HII+dust (C$_2$)  & 4 & {\bf 96} &  0 & 9 & 33 & 54 & 0  \\%Yes\\
\hline
\end{tabular}
\end{center}
\footnotesize{
Comparison of the different kind of sources approximated with a one-component (M$_1$) and a two-component model (M$_2$). We run model 2 four times setting the spectral index $\beta_c$ to the values 1.7, 2, 2.3 and 2.7 to explore the whole range indicated by~\cite{fds}. The bold face values indicate the percentage corresponding to the right input models. }
\vspace{0.5cm}
%We report also the output T-$\beta$ relationship (equation~\ref{eq:betat}) for each of the models and whether the input relationship described in Table~\ref{tab:sources} has been rightlly recovered by our analysis. For a discussion of these results see text. 
\end{table*}

\begin{table*}[!t]
\begin{center}
\space
\caption{Temperature-spectral index relationship from simulations}
\label{tab:result2}
\begin{tabular}{l | c c | c c c || c c c  }
\hline
Input Model & \multicolumn{2}{c}{Input} & \multicolumn{2}{c}{ Bayesian Output }  &  \multicolumn{1}{c}{ Input Recovery } & \multicolumn{2}{c}{ Least Square Output }  &  \multicolumn{1}{c}{ Input Recovery }\\
 & A  & $\alpha$& A(${\bf k_{ext}, \sigma_{10\%}}$) & $\alpha({\bf k_{ext}, \sigma_{10\%}}) $ & & A(${\bf \sigma_{10\%}}$) & $\alpha({\bf  \sigma_{10\%}}) $ & \\
\hline
\hline
%\multicolumn{2}{c}{} & $\beta_1,\beta_2=1.7$ & $\beta_1,\beta_2=2.0$ &$\beta_1,\beta_2= 2.3 $&\multicolumn{1}{c}{ $\beta_1,\beta_2= 2.7 $}&\\
%\hhline{~======}\\
Interstellar Medium (A$_1$) &2.0 & 0.0 & $1.8\pm0.1$ &  $0.0\pm0.1$ & True  & $1.84\pm0.03$ & $-0.8\pm0.1$& False\\ %$1.90\pm0.03$ &  $0.025\pm0.003$ &  $1.86\pm0.06$ &  $-0.06\pm0.01$\\ %Yes \\
Interstellar Medium (A$_2$) & 2.2 & -1.3 & $2.1\pm0.1$ & $-1.3\pm0.1$ & True & $2.01\pm0.03$&$-1.4\pm0.1$  & True\\ %$2.14\pm0.02$ & $-1.24\pm0.02$& $2.1\pm0.1$ &  $-1.3\pm0.1$ \\%Yes\\
Interstellar Medium (A$_3$) &  2.0 & 0.5 &$1.7\pm0.1$  & $0.4\pm0.1$ & True  & $1.83\pm0.03$ & $-0.5\pm0.1$ & False\\ %$2.14\pm0.02$ & $-1.24\pm0.02$& $2.1\pm0.1$ &  $-1.3\pm0.1$ \\%Yes\\
Cold Clump (B$_1$)  & 2.0 & 0.0 & $1.7\pm0.1$ & $-0.3\pm0.1$ & True & $1.06\pm0.01$ & $-1.67\pm0.02$ & False\\ %$1.86\pm 0.03$ & $-0.35\pm0.01$ & $1.8\pm0.1$ &  $-0.36\pm0.02$\\%Yes\\
Cold Clump (B$_2$) & 2.0 & 0.0 & $1.7\pm0.1$ & $-0.3\pm0.1$ & True & $1.42\pm0.04$& $-0.7\pm0.1$ &False\\ %$1.91\pm0.01$ & $-0.23\pm0.01$ & $1.86\pm0.03$ &  $-0.24\pm0.01$\\%Yes\\
Cold Clump (B$_3$) & 2.0 & 0.0 & $2.0\pm0.2$  & $-0.3 \pm 0.2$ & True & $0.5\pm0.1$ & $-2.3\pm0.5$ & False\\ %$2.2\pm0.1$ & $-0.27\pm0.03$&$2.0\pm0.1$ &  $-0.41\pm0.01$\\%Yes\\
Cold Clump (B$_4$) &2.0 &0.0 & $2.2\pm0.2$  & $0.0\pm0.2$ &  True & $0.8\pm0.1$ & $-1.2\pm0.3$ & False\\ % $2.1\pm0.1$ & $-0.28\pm0.04$ & $2.1\pm0.1$ &  $-0.28\pm0.08$ \\%Yes\\
HII+dust (C$_1$)   & 2.0  &0.0 & $1.9\pm0.2$ & $-0.4\pm0.7$ & True & $3.0\pm0.5$ & $-2.4\pm0.3$ & False\\ % $2.07\pm0.03$ & $-0.49\pm0.04$ & $2.0\pm0.1$ &  $-0.4\pm0.2$\\%Yes\\
HII+dust (C$_2$)   & 2.0 & 0.0 &-&-& - \\%Yes\\
\hline
\end{tabular}
\end{center}
\footnotesize{Output T$_d-\beta$ relationship (columns 4-6) obtained with our method, {based on Bayesian statistic with multi-variate Gaussian priors}, and comparison with the input (column 2-3) and with the results obtained by fitting the same set of sources with a least square method (columns 7-8). Our simulations include a 10$\%$ statistical error and a $15\%-20\%$ systematic uncertainty depending on the band. {Systematic uncertainties are treated only in the procedure based on the Bayesian method. The least square method do not allow a correct systematic error treatment, so results based on this procedure include only a 10$\%$ statistical uncertainty.}
For a discussion of these results see text. }
\vspace{0.5cm}
\end{table*}
\space

\subsection{Interstellar Medium}

The considered cases (A$_1$, A$_2$, A$_3$) are shown in Fig.~(\ref{fig:res_m1}) and reported in Tab.~(\ref{tab:1comp}). 

\begin{figure}[thb]
\begin{center}
\includegraphics[width=0.5\textwidth]{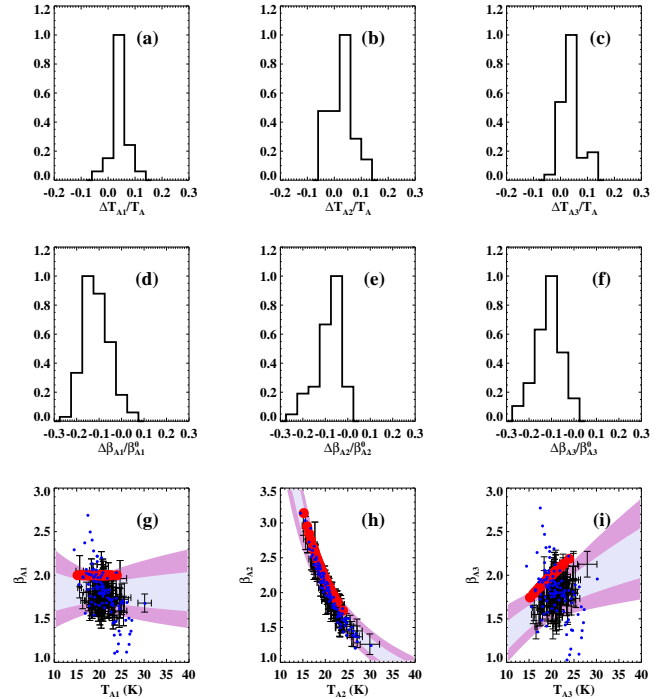}
\caption{{ Comparison between output and input temperatures and spectral indices for sources A (panels from (a) to (f)). All the histograms are normalized to 1. Panels from (g) to (i): T-$\beta$ relationships. Red circles: input values. Black dots: values recovered with the Bayesian methos, within 1-$\sigma$ bars.  Blue dots: values recovered with a least squares method. The light and dark purple contours identify the recovered correlation within 1-$\sigma$ and 2-$\sigma$ error, respectively.}}
\space
\vspace{0.5cm}
\label{fig:res_m1}
\end{center}
\end{figure}

\noindent Their SEDs are much better approximated by M$_1$ than by M$_2$, being the two-components model vary badly constrained when the Monte Carlo analysis of the calibration uncertainties is included. 
{As the plots of the relative differences show (panels from (a) to (f)), the temperatures are systematically shifted towards higher values of $\sim5-10\%$ and this correponds to a systematic spectral index shift towards lower values of $\sim10-15\%$ of the input on average. In order to understand the origin of this bias, we run five new sets of simulations of the A$_1$ model, all with $10\%$ statistical errors but with [$0\%$, $5\%$, $10\%$, $15\%$, $20\%$] calibration uncertainties, assuming all the bands to have the same systematic uncertainty. Since we do not have a more precise knowledge of the distribution of the calibration errors, we can only assume a uniform flat prior in the considered interval. We run the whole pipeline, and estimate the bias on the final temperatures ($\Delta \mathrm T/\mathrm T_0$) and spectral indices ($\Delta \beta/\beta_0$).
Results on the bias study are shown in Fig.~(\ref{fig:bias}). When a uniform wide prior is assumed (top panel), which is the case we apply on simulations and on real observations, the systematic shift increases with the calibration uncertainty, in a non-uniform way for the two parameters: the spectral index seems to be affected more than the temperature, likely because the calibration variations affect the RJ slope of the SEDs more than the peak position. Nonetheless, if we assume to have a better knowledge of the calibration uncertainty distributions, which is not the case for the observations considered in the present paper, and assume gaussian priors with standard deviations [$0\%$, $3\%$, $5\%$, $7\%$, $10\%$], the bias level is very low for both the parameters (the systematic shift is centered in few percents) and is consistent with zero.  }

\begin{figure}[!t]
\begin{center}
\includegraphics[width=0.23\textwidth]{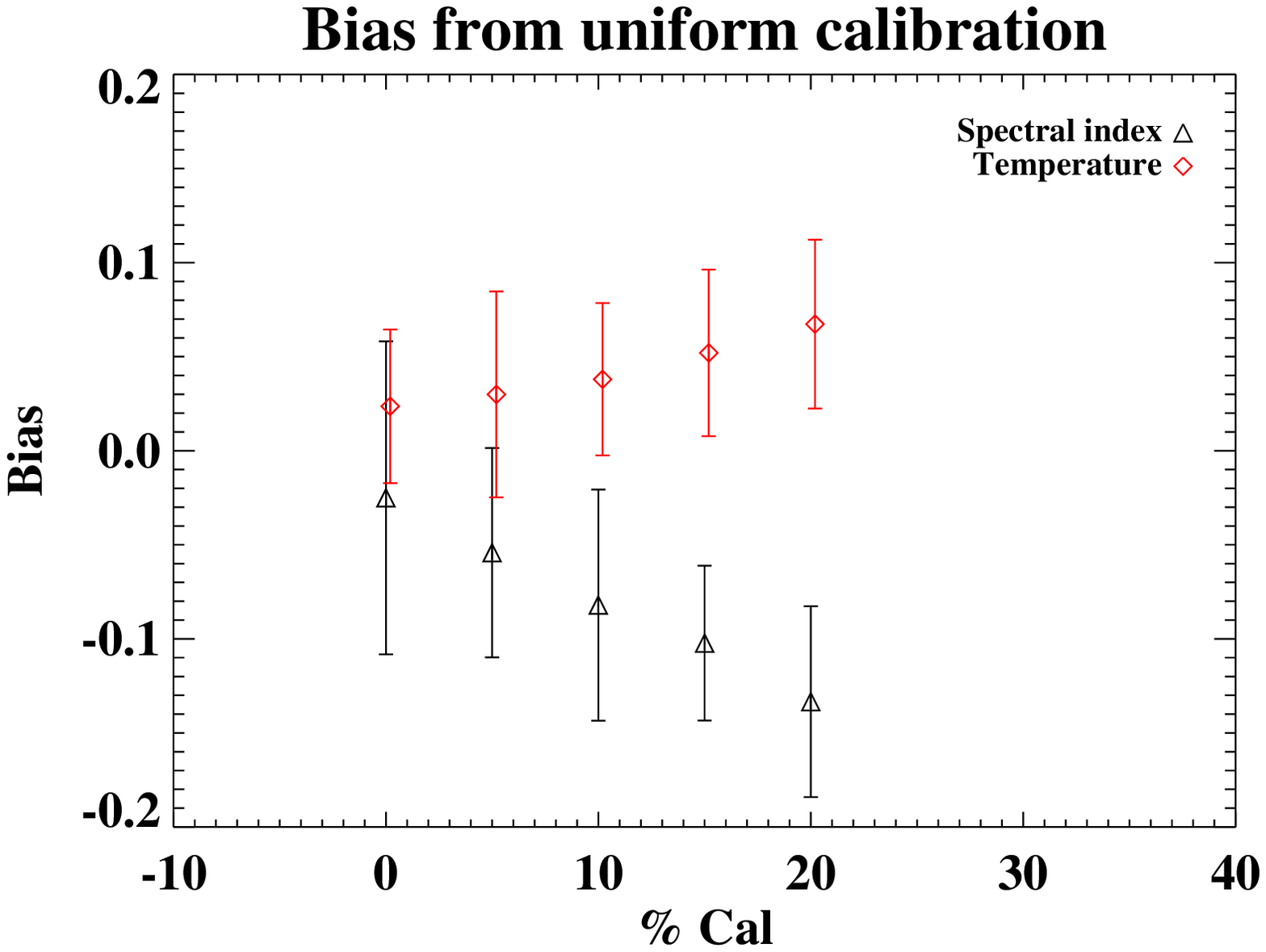}
\includegraphics[width=0.23\textwidth]{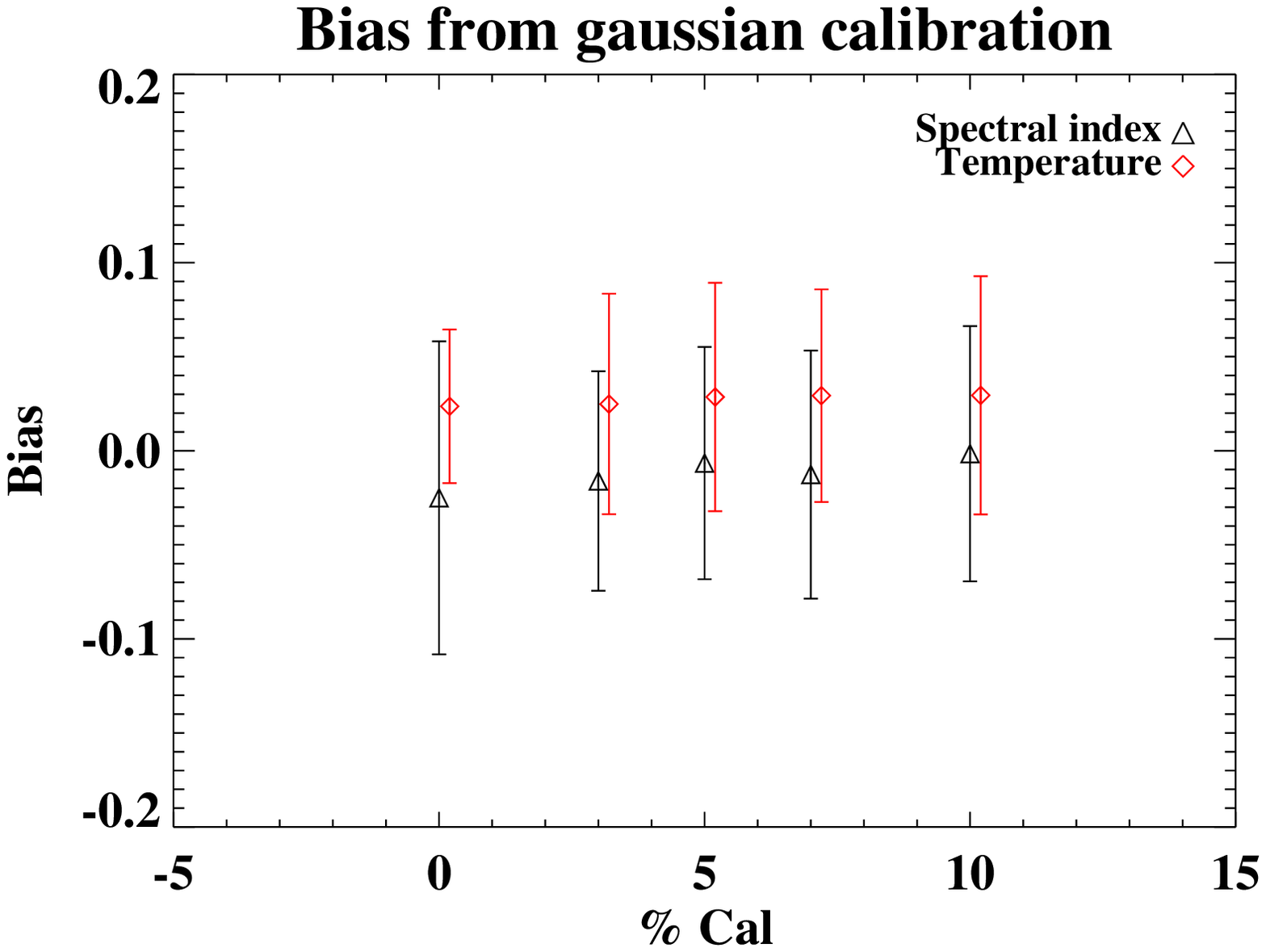}
\caption{Average bias on the final temperatures and spectral indices as a function of the calibration uncertainties. }
\label{fig:bias}
\end{center}
\end{figure}

{A good measurement of this bias would require a better knowledge of the uncertainty distribution. However, our method takes into account this effect in the estimate of the parameter uncertainties and, therefore, the input values and the recovered T$-\beta$ are consistent with the input within 1$\sigma$. 
As it will be clear in the following, all the parameter estimates in this paper are likely affected by this effect. The systematic shift is slightly more important when more components along the line of sight are assumed (cases B and C) but, again, this effect is included in the final error bars.  }

\subsection{Cold clumps}

The results of the analysis on cold clumps type of sources is reported in Tab.~(\ref{tab:1comp}) (cases B1, B2, B3 and B4) and shown in Fig.~(\ref{fig:res_case1_m3}) and~(\ref{fig:res_case2_m3}). % In Fig.~(\ref{fig:sed_2temp}) we show an example of an SED of a cold clump (left panel) and the fit performed with 

\begin{figure}[thb]
\begin{center}
\includegraphics[width=0.5\textwidth]{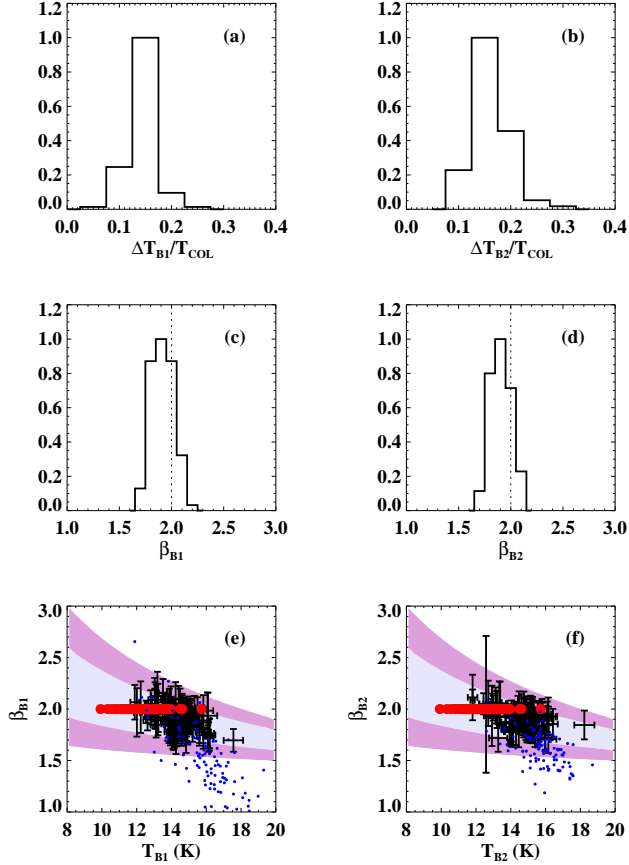}
\caption{{ Relative difference between the output temperature estimated through a one-component fit and the column temperature defined in Eq.~(\ref{eq:tcol}) for sources B$_1$ and B$_2$ (panels (a), (b)). The output spectral indices are also shown in panels (c), (d), where the dotted lines mark the input value. Panels (e) and (f): T-$\beta$ relationship. The dashed line marks the estimated output value, while the red circles show the input column temperature as a function of the input spectral index. The same parameters estimated through a least squares fit are also shown for comparison (blue dots). All the histograms are normalized to 1. The light and dark purple contours identify the recovered correlation within 1-$\sigma$ and 2-$\sigma$ error, respectively.}}
\vspace{0.5cm}
\label{fig:res_case1_m3}
\end{center}
\end{figure}

\begin{figure}[thb]
\begin{center}
\includegraphics[width=0.5\textwidth]{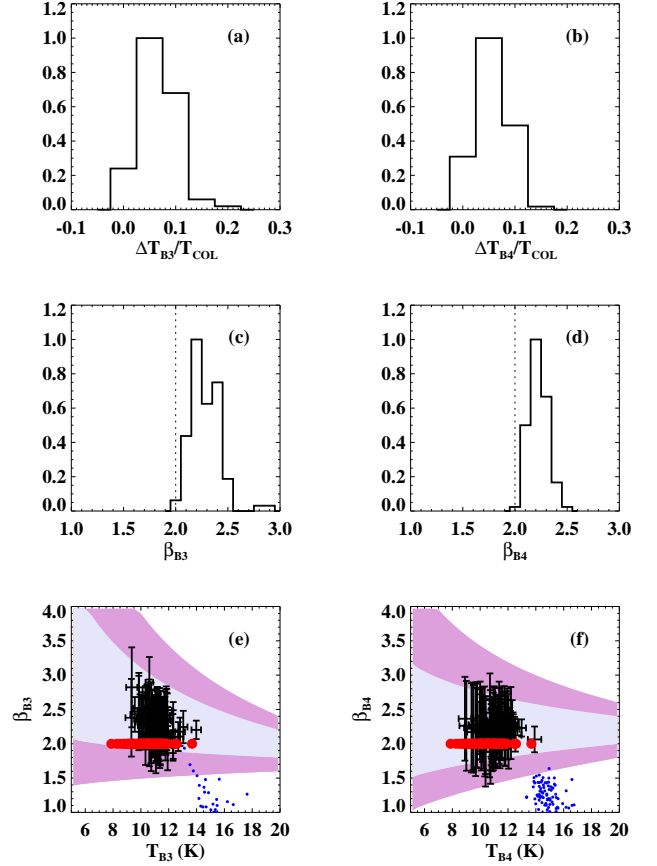}
\caption{{ Relative difference between the output temperature estimated through a one-component fit and the column temperature defined in Eq.~(\ref{eq:tcol}) for sources B$_3$ and B$_4$ (panels (a), (b)). The output spectral indices are also shown in panels (c), (d), where the dotted lines mark the input value. Panels (e) and (f): T-$\beta$ relationship. The dashed line marks the estimated output value, while the red circles show the input column temperature as a function of the input spectral index. The same parameters estimated through a least squares fit are also shown for comparison (blue dots).  All the histograms are normalized to 1. The light and dark purple contours identify the recovered correlation within 1-$\sigma$ and 2-$\sigma$ error, respectively.}}
\vspace{0.5cm}
\label{fig:res_case2_m3}
\end{center}
\end{figure}

{\noindent This is the most challenging case considered, as the SEDs are the combination of two modified black-body, having temperatures just few degrees apart. It is therefore difficult to identify the presence of two components through the comparison between M$_1$ and M$_2$, especially when the cold component has the same column density as the warm one and results, then, to be fainter (cases B1 and B2). For these reasons, the majority of sources is assigned to M1 in those cases, even if they are constituted by two emitting sources. In cases B3 and B4, since the cold component is brighter by construction, it is easier to detect the presence of two sources, even if the majority of sources is still assigned to M$_1$ and many sources have bad fit. %The 850$\mu$m point helps to better constrain the slope of the RJ part of the SED and consequently the T$--\beta$ relationship. 
Where the presence of the second component is more evident (cases B3 and B4), the one-component approximation is clearly worst and the parameters are recovered with larger uncertainties. In panels from (a) to (d) of Fig.~(\ref{fig:res_case2_m3}) we show the relative differences between the input and output column temperatures and spectral indices for these cases. Cases B1 and B2 are affected from the same systematic bias observed in the ISM simulations, due to large calibration uncertainties. Their parameters are therefore shifted of $\sim15\%$ of the original value, on average. This creates also a spurious T$-\beta$ anticorrelation which is nonetheless negligible within 3-$\sigma$s. The same is not true for cases where the cold component is denser and the column density therefore higher and comparable with the column density of the envelope. Here again the parameters, mostly the spectral indices, are affected by some bias due to the wrong modeling spectral shape. Nonetheless, the overall T$-\beta$ relationship is correct within the uncertainties and is particularly accurate for case B4, where the presence of the 850$\mu$m point better constraints the RJ slope and, therefore, the spectral index. 
The T$-\beta$ relationships are presented in both figures in panels (e) and (f). 

An application of the method on a set of cold cores on an inter-arm region of the Galactic Plane observed in the Herschel survey, is reported in Sec.~(\ref{sec:higal}). }

\subsection{HII regions and dust}

Results for HII regions-type of sources are shown in Fig.~(\ref{fig:res_m5}) and reported in Tab.~(\ref{tab:1comp}). 
\begin{figure}[th]
\begin{center}
\includegraphics[width=0.5\textwidth]{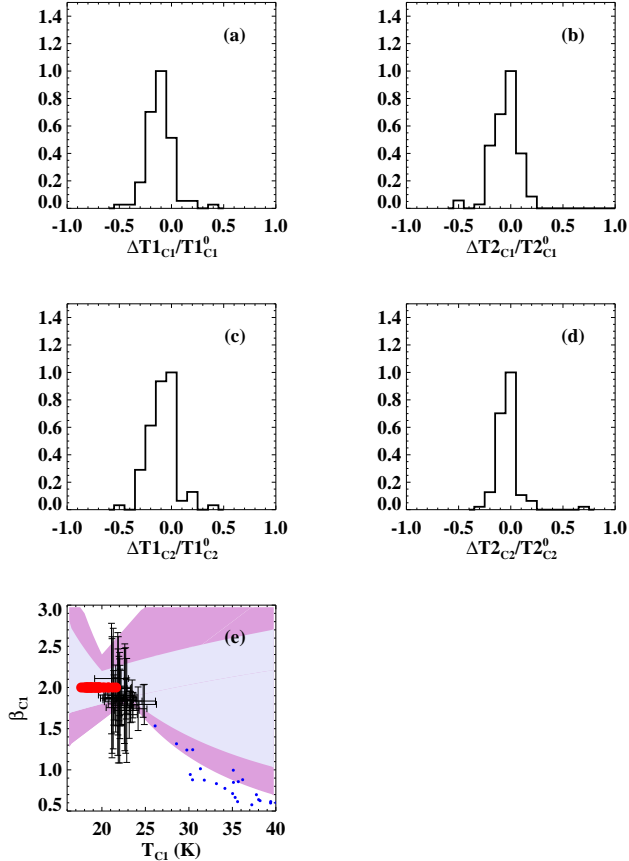}
\caption{Comparison between output and input parameters for sources C$_1$ and C$_2$. All the histograms are normalized to 1. { Panels from (a) to (d): relative difference on the cold and warm temperatures recovered with M$_2$ with respect to the input values. Panel (e): T-$\beta$ relationship from the sources C$_1$ identified as single components. The red circles mark the input column density while the light and dark purple contours identify the recovered correlation within 1-$\sigma$ and 2-$\sigma$ error, respectively. The same parameters estimated through a least squares fit are also shown for comparison (blue dots).}}
\vspace{0.5cm}
\label{fig:res_m5}
\end{center}
\end{figure}

%Since the two components have very different temperatures, it is very important to constrain the second peak. 
{\noindent In both cases the presence of two components is well detected with the model identification, especially when the 24$\mu$m flux is present to constrain the peak of the warm component (case C2). As already outlined in Sec.~\ref{sec:sources}, in this analysis we consider the 24$\mu$m flux to be dominated by BG. For this kind of sources, M$_2$ fits better than M$_1$ in 100\% of cases. When excluding the MIPS point (case C1) and we sample the SEDs from 70$\mu$m down to 500$\mu$m, $\sim70\%$ of cases are properly classified. Panels from (a) to (d) show the correct recovery of the two component temperatures (ISM and HII) by fitting the SEDs with M2. 
They are recovered with a maximum relative error of 50$\%$ in the worst cases (panels from (a) to (d)). Panel (e) shows the T$-\beta$ relationship for sources C1, by fitting the SEDs from 70$\mu$m to 500$\mu$m with a modified single black-body. The 70$\mu$m flux might still be affected from the warm component emission and this creates an average shift towards warmer temperatures in the parameters recovery.
Despite a slight anticorrelation is detected on the data points, it results to be negligible when the calibration uncertainties are taken into account. }

\section{Starless cores in the Herschel/Hi-GAL survey}\label{sec:higal}

After having tested the validity of the method on simulated datasets, we apply it on a set of starless cores detected in the Herschel Hi-GAL survey~\citep{Molinari10a, Molinari10b} in an interarm region of the Galactic plane. 
{Hi-GAL is one of the Herschel open time key projects to map the entire Galactic plane in $2^\circ\times2^\circ$ tiles with PACS and SPIRE in parallel mode, in the bands 70, 160, 250, 350, 500 $\mu$m.  Once the survey will be complete, the sky coverage will be $0^\circ<\ell<360^\circ$ and $-1^\circ<\ell<1^\circ$. The dataset, due to the spectral range, sensitivity and sky coverage, is particularly sensitivity to the very early stages of high-mass star formation. 
In this paper, we focus on one of the Science Demonstration Phase (SDP) fields of the survey: a $2^\circ\times2^\circ$ tile centered on ($\ell$, $b$) = (59$^\circ$, 0$^\circ$). This field has been observed in November 2009 and, since then, it has been largely studied and used to test the analysis algorithms.} 
{The tile covers an interarm region, since the line of sight is tangent to the Sagittarius arm, at an heliocentric distance between 2 and 7 kpc, where most of the sources are located.
The overall star formation activity is mostly dominated by the Vulpecula OB association (see for example,~\cite{Billot10}), a molecular complex with triggered star forming activity taking place. Nonetheless, the Star Formation Rate (SFR) in that region, estimated from the 70$\mu$m colors of the observed Young Stellar Objects (YSOs) is low ($2.6\times10^{-6}$ M$_\odot/$yr,~\cite{Veneziani13}). Since the background emission is also weak, the source detection is very accurate down to very low signal-to-noise thresholds  and this makes it a favorable set to study molecular cold clumps in a prestellar evolutionary stage. }
  
 \subsection{Observations and data analysis}
  
{In this context, we use the data from the $Spitzer$ legacy survey MIPSGAL at 24$\mu$m~\cite{Carey09} and the Herschel-HiGAL data at 70, 160, 250, 350, 500~$\mu$m. Kinematic sun distances are also estimated from CO observations~\cite{Russeil11}.}

{Herschel maps have been produced by means of the ROMAGAL algorithm, based on a Generalized Least Square technique, after a careful preprocessing in which glitches and systematic effects present in the data have been removed. For further information about the pipeline followed from raw data up to sky maps delivery, we refer the reader to~\cite{Traficante11}.}
{The source detection and extraction has been performed by means of the CuTEX algorithm~\cite{Molinari11} which double-differentiates the sky image and study the curvature variations above a given threshold. The identified source profiles are then fitted with a 2D Gaussian plus an underlying inclined planar plateau. The Herschel catalog is then produced in the following way:  when a source is found in the longest wavelength map (500$\mu$m), we look for an association in the following band (350$\mu$m) in a radius as large as the FWHM of the largest beam between the two. If a source is found, then we proceed to the association with the following band (250$\mu$m) and so on up to the 70$\mu$m map. If more than a source is found in the radius, we associate the nearest. Fluxes corrections are applied both in the PACS and SPIRE bands in order to take into account of the non perfectly Gaussian shape of the beam. }
{By merging the Herschel catalog with the MIPS catalog by means of the same procedure described above, we have an overall catalog of the entire region from 24 to 500 $\mu$m.}

{The procedure to select and identify the starless molecular clumps is summarized in the following:  

\begin{enumerate}
\item{the sources are not detected either in the 70$\mu$m nor in the 24$\mu$m band in order to exclude the presence of a forming star inside the clump; }
\item{the sources are detected in all the four remaining bands, from 160$\mu$m to 500$\mu$m. This makes us sure not to include spurious detections and to have a better characterization of the SEDs;}
\item{the sources are gravitationally bound. As in~\cite{Veneziani13}, we identify as bounded objects the sources with mass M $\geqslant$ 0.5 M$_{BE}$ where M$_{BE}$ is the Bonnor-Ebert mass. A Bonnor-Ebert sphere is an isothermal sphere at hydrostatic equilibrium. Therefore, 
in absence of internal turbulence and assuming thermal pressure, the BE mass is a good approximation of the virial mass. }

\end{enumerate}  
}
{The absence of turbulence is a necessary approximation due to the fact that no spectroscopic data are available on this sample of sources. We cannot then check the internal motions occurring in the clumps and see if there are supporting mechanisms other than thermal pressure.   }
After all these criteria have been applied we have a total of 103 sources. Since only four fluxes are available, we assign an upper limit of 0.2 Jy to the 70$\mu$m band, in order to perform the model identification. With Hi-GAL sensitivity and coverage, 0.2 Jy is the minimum flux detected in the PACS blue band even at a 15 kpc distance~\citep{Veneziani13}. 5 sources are excluded through the model identification, leaving us with a sample of 91 sources. {The majority of those objects have radius $\delta>0.1$ pc, confirming that they are essentially clumps. }

\subsection{SEDs fitting and dust parameter estimate}

{We associate a conservative error bar of 30$\%$ of the fluxes to the SEDs from 160$\mu$m to 500$\mu$m. This error takes into account the gaussian random noise, the source multiplicity, the background fluctuations and the overall uncertainty on fluxes recovery of the extraction and photometry algorithms. The calibration errors included are PACS and SPIRE uncertainties on extended emission, the same we considered in our simulation, i.e. 20$\%$ for the 160$\mu$m PACS band and 15$\%$ for the 250, 350 and 500$\mu$m SPIRE bands. 

The Bayesian fit with multi-variate Gaussian priors and following marginalization on the calibration uncertainties are then run to estimate the SEDs parameters. After a first MCMC run with uniform wide priors, we fix the center of the multi-variate Gaussian priors in the point [T$_0$, $\log(\epsilon_0)$, $\beta_0$] = [12, -9, 2]. As already described, even if the full sample is collected in the same observation, the covariances are estimated source by source since they depend on random fluctuations, SED sampling and overlap of other sources along the line of sight, assuming that all the other sources of error, i.e. systematic effects, are the same for all the objects in the same observation.
In each covariance matrix, the standard deviations are enlarged by ten times in order to make sure not to constrain the final parameters with a too tight distribution but keeping the same correlations. The values are nevertheless well described by the ones obtained from the simulation run of cold cores and reported in Tab.~(\ref{tab:priors2}).}

{An example of the SEDs with calibration and statistical uncertainties and their best fits, is shown in Fig.~\ref{fig:sed}.}

\begin{figure}[!t]
\begin{center}
\includegraphics[width=0.5\textwidth]{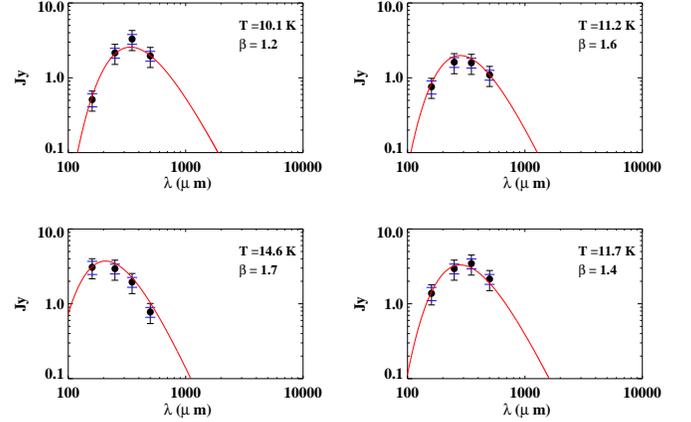}
\caption{Example of SEDs of four cold clumps at different temperatures and spectral indices values. The black symbols mark the fluxes with the associated 30$\%$ error bars. The blue dashes indicate the calibration uncertainty. }
\label{fig:sed}
\end{center}
\end{figure}

{The final temperature and spectral index distributions, as well as the T-$\beta$ relationship recovered with our bayesian method (BM), are shown in Fig.~(\ref{fig:dati}) as black dots with the associated error bars. Their median values are $\left<T_{\ell59}\right> = 11.8$ K and  $\left<\beta_{\ell59}\right> = 1.6$.
The relationship between the temperature and the spectral index  has parameter values $A_{\ell59}^{BM} = 1.91\pm0.29$ and $\alpha_{\ell59}^{BM}=0.28\pm0.29$. 
In order to compare the results of our Bayesian method with a technique based on least squares minimization, we fit the SEDs by associating to the fluxes only a statistical 30$\%$ error bar, without taking into account the calibration uncertainties. The temperature and spectral indices values are shown in Fig.~(\ref{fig:dati}) as red dots. 
Results from the least square fitting show a spurious clear anticorrelation: $A_{\ell59}^{LS} = 0.8\pm0.3$ and $\alpha_{\ell59}^{LS} = -1.3\pm0.5$ which is, nonetheless, consistent with no anticorrelation within a 3$\sigma$ range.}

{As already described in the Introduction an increase of the emissivity spectral index in cold environments had been observed both in lab experiments~\citep{Meny07} and in real observation~\citep{Dupac03,Desert08,Veneziani10,Paradis10,Bracco11}. It can be explained in two ways: the spectral index is temperature dependent, as suggested from~\cite{Meny07} from lab experiments on amorphous silicate-based dust grains, or the spectral index increases as a consequence of grain aggregation in cold and dense environments~\citep[see for example][]{Ormel11}. In the observed Herschel cold cores, we do not detect a significant increase of the spectral index with respect to the ISM, since $\beta$ values are spread between 1 and 2.5 (Fig.~\ref{fig:dati}, middle panel) and, when the T-$\beta$ trend is studied, points are scattered on the T-$\beta$ plane excluding the possibility of a correlation (Fig.~\ref{fig:dati}, right panel). The absence of a correlation and, more in general, the lack of an increase of the spectral index with respect to the diffuse ISM, can be explained considering that, as pointed out several times during the paper, this analysis is very sensitive to systematic effects which might range from instrumental issues other than calibration, (i.e. photometric extraction, catalog band-merging), to multiple overlap of sources along the line of sight. A possible physical explanation of our findings might be that we are observing cores of different ages and this reflects the spread of the spectral indices. As~\cite{Ossenkopf94} shows, when silicate dust grains are covered by a thick ice mantle, their emission becomes independent of the optical properties of the material underlying the mantle, as to mask the T-$\beta$ anticorrelation. Therefore, depending on the mantle thickness, i.e. the timescale of the core, the optical properties of the grains and their spectral indices change. Moreover, as~\cite{Ormel09} shows, depending on the cold core timescale, grains might or might not have the time to aggregate and this also can explain the spread of the spectral index values.  }

}

\begin{figure*}[!t]
\begin{center}
\includegraphics[width=0.9\textwidth]{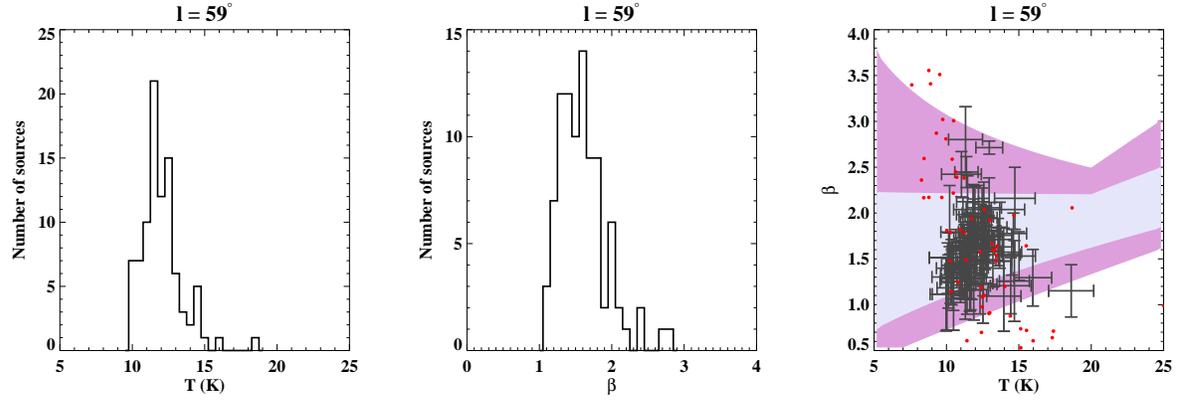}
\caption{Physical parameters of the starless clumps in the Herschel/Hi-GAL ($\ell$, $b$) = ($59^\circ$, 0) field. First and second panels show the temperature and spectral index distributions, respectively. The third panel shows the T-$\beta$ distribution. The light and dark purple contours identify the recovered correlation within 1-$\sigma$ and 2-$\sigma$ error, respectively. The red dots mark the temperatures and spectral indices obtained by fitting the SEDs with a least square method, for comparison.  }
\label{fig:dati}
\end{center}
\end{figure*}

\section{Conclusions}\label{sec:conclusions}

We have presented a method, based on Bayesian statistics, to fit for intrinsically correlated parameters and look for the underlying correlation law. This method is different from the ones previously used for similar analysis, as we make use of Bayesian statistics, which fully sample the parameter space taking into account the covariance among different parameters, together with a proper treatment of systematic errors by means of a Monte Carlo procedure. 
We apply the method to estimate the physical properties and the relationship between the temperature and the spectral index of a set of simulated astrophysical sources detectable in the Herschel PACS and SPIRE bands (between 70$\mu$m and 500$\mu$m), including also Planck-HFI (850$\mu$m), IRAS (100$\mic$) and MIPS (24$\mu$m) bands, in order to better constrain the colder and warmer components. The sources are chosen over a wide range of temperatures and compositions in order to test the efficiency of the method with different samplings of the SEDs. For this purpose, we consider the cases of a one-component ISM (with T$_d$ and $\beta$ either correlated, case A2 and A3, or uncorrelated, case A1), and of two-component temperature sources both warm and cold. The case of two-component sources is particularly interesting because, as shown in~\cite{Shetty09a,Shetty09b}, the combination of more than one source along the line of sight generates a spurious anticorrelation by itself. 

The most relevant results are:\\

\begin{itemize}

{ 
\item {The input physical values and their correlations are well recovered in the Herschel bands when we observe an ISM like sample both without and with correlation (cases A1, A2 and A3, respectively); }

\item{when two sources just few degrees apart are combined, as cold clumps, it is difficult to identify the underlying model especially when the core and the envelope have the same column density (cases B1 and B2) and the core emission is therefore much fainter. If the core is thicker it is easier to detect (cases B3 and B4) and, in this case, the presence of the 850$\mu$m flux (B4) helps to better constrain the RJ part of the SEDs and, therefore, the physical parameters and their correlations. 
 Even if a spurious anticorrelation is detected in the cold clumps, when taking into account the calibration uncertainties, the detected anticorrelation is negligible within 3$\sigma$.  }
 
 \item{when the source is a combination of two components several degrees apart, as for HII regions, the underlying model is correctly identified, especially when the 24$\mu$m flux is present. Also in this case, a slight anticorrelation is measured in the dust component but the detection is neglibile when taking into account calibration uncertainties.  }
}
\end{itemize}

{ 
The method has been applied to a sample of starless clumps in a $2^\circ\times2^\circ$ field centered on ($\ell$, $b$) = ($59^\circ$, 0) observed in the  Herschel/Hi-GAL survey.  Being an inter-arm region of the Galactic plane, it is not not very dense with sources and the star formation activity is therefore very low. The average temperature and spectral index of the sample are $\sim11.8$ K and 1.6, respectively, as expected for cold sources. 
We do not detect a T$-\beta$ correlation within the uncertainties.  

}

\section{Acknowledgments}
%The authors thank Brandon Kelly, Rahul Shetty and Alyssa Goodman for sharing their results of their analysis on SED fitting using also the
%MCMC method prior publication. 
MV acknowledges ASI support via contract I/038/080/0. The authors thank the anonymous referee for helpful suggestions which greatly improved the quality of the paper.

\newpage

% \appendix

%\vspace{0.5cm}


\newpage

\begin{thebibliography}{} 

\bibitem[Billot et al.(2010)]{Billot10} Billot, N., Noriega-Crespo, A., Carey, S. et al.,
2010, ApJ, 712, 797 
\bibitem[Bracco et al.(2011)]{Bracco11} Bracco, A., Cooray, A., Veneziani, M. et al.,
2011, MNRAS, 412, 1151
\bibitem[Carey et al.(2009)]{Carey09} Carey, S. J., Noriega-Crespo, A., Mizuno, D. R, et al.,  
2009, PASP, 121, 76
\bibitem[Coupeaud et al.(2011)]{coupeaud2011} Coupeaud, A.,  Demyk, K., Meny, C. et al.,
2011, A$\&$A, 535, A124
\bibitem[D\'esert et al.(1990)]{Desert90} D\'esert, F.-X., Boulanger, F., $\&$ Puget, J. L.,
1990, A$\&$A, 237, 215
\bibitem[D\'esert et al.(2008)]{Desert08} D\'esert, F.-X., Mac\'ias-P\'erez, J. F., Mayet, F., et al.,
2008, A$\&$A, 481, 411
\bibitem[Doty $\&$ Palotti(2002)]{Doty02} Doty, S. D.,  Palotti, M. L.,
2002, MNRAS, 335, 993 
\bibitem[Draine $\&$ Li(2007)]{Draine07} Draine, B. T. $\&$ Li, A.,
2007, ApJ, 657, 810
\bibitem[Dupac et al.(2003)]{Dupac03} Dupac, X., Bernard, J.-Ph., Boudet, N., et al.,
2003, A$\&$A, 404, L11
\bibitem[Finkbeiner et al.(1999)]{fds} Finkbeiner, D.~P., Davis, M., \& Schlegel, D.~J., 
1999, \apj, 524, 867
\bibitem[Griffin et al.(2010)]{Griffin10} Griffin, M. J., Abergel, A., Abreu, A., et al.,
2010, A$\&$A, 518, L3
\bibitem[Juvela $\&$  Ysard(2012a)]{Juvela12a} Juvela, M., Ysard, N.,
2012, A$\&$A, 539, 71
\bibitem[Juvela $\&$  Ysard(2012b)]{Juvela12b} Juvela, M., Ysard, N.,
2012, A$\&$A, 541, 33
\bibitem[Kelly et al.(2011)]{Kelly12} Kelly, B.C., Shetty, R., Stutz, A. M., et al.,
2012, ApJ, 752, 55
\bibitem[Lewis $\&$ Bridle(2002)]{Lewis02} Lewis, A., $\&$ Bridle, S.,
2002, Phys. Rev., 66, 103511
\bibitem[M\'eny et al.(2007)]{Meny07} M\'eny, C., Gromov, V., Boudet, N., et al.,
2007, A$\&$A, 468, 171
\bibitem[Miville-Desch\^enes $\&$ Lagache(2005)]{MamD05} Miville-Desch\^enes, M. A., $\&$ Lagache, G.,
2005, ApJS, 157, 302
\bibitem[Molinari et al.(2010a)]{Molinari10a} Molinari, S., Swinyard, B., Bally, J., et al.,
2010, PASP, 122, 314
\bibitem[Molinari et al.(2010b)]{Molinari10b} Molinari, S., Swinyard, B., Bally, J., et al.,
2010, A$\&$A, 518L, 100
\bibitem[Molinari et al.(2011)]{Molinari11} Molinari, S., Schisano, E., Faustini, F., et al.,
2011, A$\&$A, 530, A133
\bibitem[Ormel et al.(2009)]{Ormel09} Ormel, C., W.,  Paszun, D., Dominik, C., et al., 
2009, A$\&$A, 502, 845
\bibitem[Ormel et al.(2011)]{Ormel11} Ormel, C., W.,  Min, M., Tielens,  A., G., G., M., et al., 
2011, A$\&$A, 532, A43
\bibitem[Ossenkopf $\&$ Henning(1994)]{Ossenkopf94} Ossenkopf, V., $\&$, Henning, T.,
1994, A$\&$A, 291, 943
\bibitem[Paladini et al.(2012)]{Paladini12}Paladini, R., Umana, G., Veneziani, M., et al.,
2012, ApJ, 760, 149
\bibitem[Paradis et al.(2010)]{Paradis10} Paradis, D., Veneziani, M., Noriega-Crespo, A., et al.,
2010, A$\&$A, 520, L8
\bibitem[Pilbratt et al.(2010)]{Pilbratt10} Pilbratt, G., Riedinger, J. R., Passvogel, T., et al.,
2010, A$\&$A, 518, L1
\bibitem[Planck Collaboration(2011a)]{planck_mission} Planck Collaboration,
2011, A$\&$A, 536, A1
\bibitem[Planck Collaboration(2011b)]{planck_cold_cores} Planck Collaboration, 
2011, A$\&$A submitted, arXiv: 1101.2035
\bibitem[Poglitsch et al.(2010)]{Poglitsch10} Poglitsch, A., Waelkens, A., Geis, N., et al.,
2010, A$\&$A, 518, L2
\bibitem[Russeil et al.(2011)]{Russeil11} Russeil, D., Pestalozzi, M., Mottram, J. C., et al.,
2011, A$\&$A, 526, A151
\bibitem[Salgado et al.(2012)]{salgado2012} Salgado, F., Berne, O., Adams, J. D., et al.,
2012, ApJ, 749, L21
\bibitem[Shetty et al.(2009a)]{Shetty09a} Shetty, R., Kauffmann, J., Schnee, S., et al.,
2009, ApJ, 696, 676
\bibitem[Shetty et al.(2009b)]{Shetty09b} Shetty, R., Kauffmann, J., Schnee, S., et al.,
2009, ApJ, 696, 2234
\bibitem[Traficante et al.(2011)]{Traficante11}  TraÞcante, A., Calzoletti, L., Veneziani, M., et al. 2011, 
MNRAS, 416, 2932 
\bibitem[Veneziani et al.(2010)]{Veneziani10}Veneziani, M., Ade, P. A. R., Bock, J. J., et al.
2010, ApJ, 713, 959
\bibitem[Veneziani et al.(2013)]{Veneziani13} Veneziani, M., Elia, D., Noriega-Crespo, A., et al.,
2013, A$\&$A, 549, A130

\end{thebibliography}
\end{document}